\documentclass[final,times,3p,lefttitle]{elsarticle}
\usepackage{lineno}
\usepackage[svgnames,dvipsnames]{xcolor}
\usepackage{colortbl,array}
\usepackage{amsmath,amssymb,amsfonts}
\usepackage{graphicx}
\usepackage{multirow}
\usepackage{here}
\usepackage{bm}
\usepackage{natbib}
\usepackage{url}
\usepackage{textcomp,nicefrac}

\journal{arXiv}

\usepackage{docmute}

\begin{document}

\begin{frontmatter}

\title{\LARGE Imaging and Spectral Performance of \\ a Wide-gap CdTe Double-Sided Strip Detector\vspace{-5pt}}

% Group authors per affiliation:
\author[ssl,ut,kavli]{Shunsaku Nagasawa\corref{cor1}}
\author[ut,kavli]{Takahiro Minami}
\author[isas,kavli]{Shin Watanabe}
\author[kavli,ut]{Tadayuki Takahashi}

\address[ssl]{Space Sciences Laboratory, University of California, Berkeley, 7 Gauss Way, Berkeley, CA 94720, USA}
\address[ut]{Department of Physics, The University of Tokyo, 7-3-1 Hongo, Bunkyo, Tokyo 113-0033, Japan}
\address[kavli]{Kavli Institute for the Physics and Mathematics of the Universe (Kavli IPMU, WPI), The University of Tokyo, 5-1-5 Kashiwanoha, Kashiwa, Chiba 277-8583, Japan}
\address[isas]{Institute of Space and Astronautical Science, Japan Aerospace Exploration Agency (ISAS/JAXA), 3-1-1 Yoshinodai, Chuo-ku, Sagamihara, Kanagawa 252-5210, Japan}

\cortext[cor1]{Email Address: nshunsaku@berkeley.edu}

\begin{abstract}
The fourth flight of the Focusing Optics X-ray Solar Imager sounding rocket experiment (FOXSI-4) aimed to achieve the first imaging spectroscopic observations of mid-to-large class ($\geq$ GOES C5 class) solar flares, in contrast to the previous three flights that targeted relatively quiet regions of the Sun.
To meet the emerging requirements for hard X-ray focal plane detectors for providing simultaneous diagnostics of spectrally ($<1$ keV FWHM) and spatially ($<50~\mathrm{\mu m}$) separated coronal and chromospheric emissions from solar flares, we developed a new strip-configuration detector called the wide-gap CdTe semiconductor double-sided strip detector (CdTe-DSD).

The wide-gap CdTe-DSD employs a unique design principle to enhance position resolution. This enhancement is realized by expanding the gaps between electrodes to induce charge-sharing across adjacent strip electrodes and using this sharing energy information for position reconstruction to a level finer than the strip-pitch.
However, the detector response becomes complex and requires consideration of various factors, such as the charge loss due to wider gaps and the dependence on the depth of photon interaction. 
Thus, we developed an energy reconstruction method that fully leverages the energy information between adjacent strips and from both the cathode and anode sides, achieving an energy resolution of 0.75 keV (FWHM) at 14 keV.
Furthermore, we conducted an X-ray scanning experiment using a synchrotron beam at Spring-8 to evaluate the detector response with a fine scale of 10 $\mathrm{\mu m}$. Based on these results, we established a sub-strip position reconstruction method, demonstrating that X-rays interacting at the center of the gap can be determined with an accuracy of 20 $\mathrm{\mu m}$, and even those at the strip center can be determined with an accuracy of 50 $\mathrm{\mu m}$.

\end{abstract}

\begin{keyword}
CdTe,
Charge Sharing,
Strip Detector, 
X-rays,
Solar Flares
\end{keyword}

\end{frontmatter}

\section{Introduction}
Imaging spectroscopy in hard X-rays is widely utilized across astronomy, industry, and medical imaging. We have been developing double-sided strip detectors (DSDs) based on cadmium telluride diodes (CdTe), referred to as CdTe-DSDs, which offer superior energy resolution due to the implementation of a Schottky junction on the CdTe surface \cite{takahashi1999high, takahashi2001recent, ishikawa2016fine}.
The DSD configuration provides advantages over pixel-type detectors by allowing signal measurements from both anode and cathode strips. This double-sided readout leverages the differing mobility-lifetime products ($\mu \tau$) of holes and electrons in CdTe, enabling depth-of-interaction (DoI) estimation and improving energy and position resolution \cite{takahashi2001recent, furukawa2019development, minami20232mmthick}.
Additionally, we can utilize charge sharing, where the charge cloud may extend across multiple strips based on the interaction position of the photon. Leveraging the signal ratio between adjacent strips allows for enhanced position resolution, surpassing the limitations imposed by the strip-pitch \cite{koenig2013charge, furukawa2020imaging}.

The Focusing Optics X-ray Solar Imager (FOXSI) sounding rocket experiment \cite{krucker2009focusing} is the first Solar-dedicated mission to realize the direct imaging method combined with Wolter-I hard X-ray focusing optics and fine-pitch focal plane detectors.
The fourth flight, FOXSI-4, was conducted as NASA's first ``Flare Campaign" \cite{savage2022first, glesener2022high} and was launched at the same time as solar flares occur to achieve the first direct imaging observation of mid-to-large class ($\geq$ GOES C5 class) solar flares, while the past three flights targeted the relatively quiet regions of the Sun \cite{krucker2009focusing, glesener2016foxsi, christe2016foxsi, musset2019ghost}.

For the FOXSI-4 hard X-ray focal plane detectors, covering the energy range of 4 keV to 20 keV, an energy resolution of 1 keV and a position resolution of 50~$\mathrm{\mu m}$ are simultaneously required to provide simultaneous diagnostics of the spectrally and spatially separated coronal and chromospheric emission of solar flares in hard X-rays \cite{krucker2008hard}.
Furthermore, the focusing optics are being improved to achieve an angular resolution of $\lesssim 2$ arcsec (FWHM), which corresponds to a position resolution of $\lesssim 20~\mathrm{\mu m}$ at a focal length of 2 m \cite{Camilo2021, yoshida2024development, ampuku2024development}.
Therefore, to fully exploit the enhanced optics performance, a position resolution less than $20~\mathrm{\mu m}$ is desired.

For the third flight, FOXSI-3 \cite{furukawa2019development, furukawa2020imaging}, we developed the 60 $\mathrm{\mu m}$ strip-pitch CdTe-DSD (50~$\mathrm{\mu m}$ strip and 10~$\mathrm{\mu m}$ gap width).
We demonstrated that by modeling the charge propagation based on the information of the DoI and sharing energies between adjacent strips, a position resolution finer than the strip-pitch can be achieved.
However, the percentage of charge-sharing events for 20 keV is limited to 25\% on the cathode side \cite{furukawa2020imaging}. 
To fully utilize charge-sharing and achieve sub-strip position resolution, it is necessary to enhance the occurrence of charge-sharing events.

Although it is difficult to reduce a strip-pitch further to ensure a certain yield level with current technology, it is expected to enhance charge-sharing events by widening the gap between strips. We have previously demonstrated this concept by developing a one-sided wide-gap strip detector \cite{nagasawa2023wide}.
By adjusting the gap width, we confirmed that the occurrence of double-strip events, where signals are detected on two adjacent strips, can be increased to approximately 50\% on the cathode side \cite{nagasawa2023wide}.

In this paper, we present a novel concept of CdTe-DSD, termed ``Wide-gap CdTe-DSD", designed for the hard X-ray focal plane detectors of FOXSI-4. In this design, the gap between adjacent strips is much wider than the width of each strip.
In Section \ref{sec:det_conf}, we describe the configuration of the wide-gap CdTe-DSD. In Section \ref{sec:spec}, we evaluated the spectral performance through uniform X-ray irradiation tests using radioisotope (RI) sources. In Section \ref{sec:spring8}, we present results from synchrotron beam scanning tests to evaluate the charge-sharing properties for sub-strip position reconstruction.
Based on these results, we developed a new energy-reconstruction method to compensate for charge loss caused by the wider gap and DoI effects, enhancing the detection efficiency while maintaining the spectrum performance. Additionally, we developed a sub-strip position reconstruction method based on the relationship between the incident position and the detected energies on adjacent strips.

\section{Detector configuration}\label{sec:det_conf}
\subsection{Wide-gap CdTe-DSD}
\begin{figure}[htb]
  \begin{center}
  \includegraphics[width=1.\hsize]{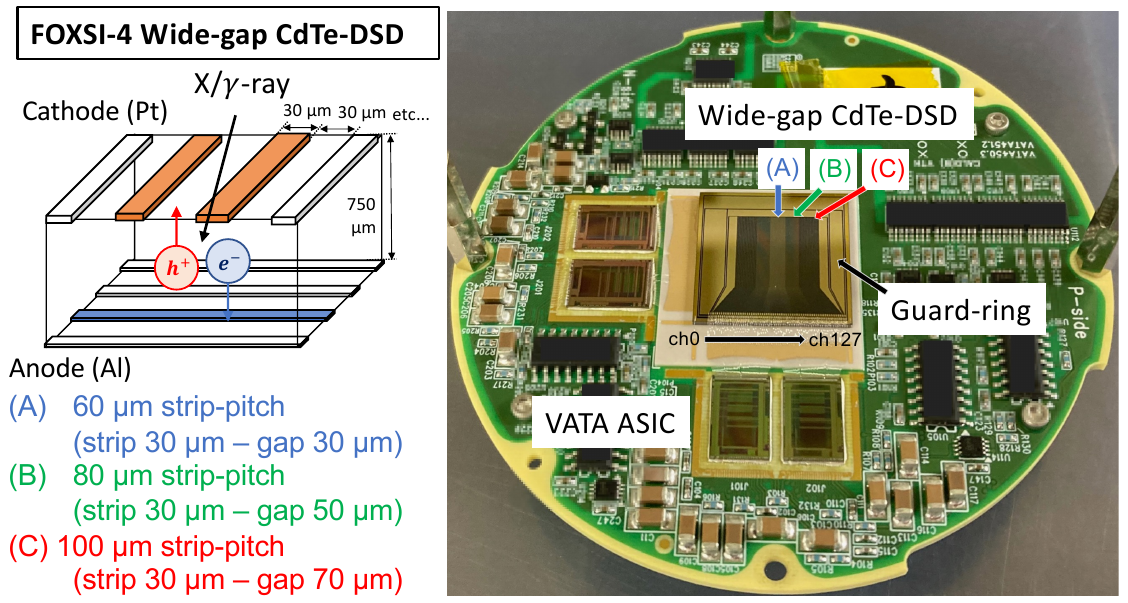}
  \caption{Schematic diagram and picture of the Front End Card (FEC) of the wide-gap CdTe-DSD for FOXSI-4.}
  \label{foxsi4_dsd}
  \end{center}
\end{figure}

Fig.\,\ref{foxsi4_dsd} shows a picture of the wide-gap CdTe-DSD for FOXSI-4. The detector has a thickness of 750 $\mathrm{\mu m}$ with a sensitive area of 9.92$\times$9.92 $\mathrm{mm^2}$ and has 128 strip electrodes orthogonally placed on both the anode (Al) and cathode (Pt) sides.
The width of the strips is 30 $\mathrm{\mu m}$ in common, but the width of gaps is varied from 30 $\mathrm{\mu m}$ to 70 $\mathrm{\mu m}$ from the center to the edge of the detector, as summarized in Table \ref{ch_map}.
Each cathode and anode side is divided into three regions with different strip pitches, resulting in a total of nine regions across the detector (three strip pitches on each side $\times$ two sides).
This design allows us to study how varying gap widths affect the charge-sharing properties on both the cathode and anode sides.
The strip electrodes are surrounded by guard-ring electrodes, which reduce leakage current through the surface of the CdTe and improve the energy resolution significantly \citep{nakazawa2004improvement}.

\begin{table}[htb]
  \centering
  \caption{Strip-pitch and strip/gap width of FOXSI-4 wide-gap CdTe-DSD}
  \label{ch_map}
  \small
  \begin{tabular}{l|l}
  \hline
  0 ch -- 3 ch & Guard-ring electrodes \\
  4 ch -- 27 ch & (C) 100~$\mathrm{\mu m}$ strip-pitch (strip 30~$\mathrm{\mu m}$ -- gap 70~$\mathrm{\mu m}$)\\
  28 ch -- 47 ch & (B) 80~$\mathrm{\mu m}$ strip-pitch ~~(strip 30~$\mathrm{\mu m}$ -- gap 50~$\mathrm{\mu m}$)\\
  48 ch -- 63 ch & (A) 60~$\mathrm{\mu m}$ strip-pitch ~~(strip 30~$\mathrm{\mu m}$ -- gap 30~$\mathrm{\mu m}$)\\
  80 ch -- 99 ch & (B) 80~$\mathrm{\mu m}$ strip-pitch ~~(strip 30~$\mathrm{\mu m}$ -- gap 50~$\mathrm{\mu m}$)\\
  100 ch -- 123 ch & (C) 100~$\mathrm{\mu m}$ strip-pitch (strip 30~$\mathrm{\mu m}$ -- gap 70~$\mathrm{\mu m}$)\\
  124 ch -- 127 ch & Guard-ring electrodes  \\
  \hline
  \end{tabular}
\end{table}

The detector is mounted on an FEC (Front End Card) board and connected to a DIO (Digital Input/Output) board and a power supply DC-DC board. 
The detector is connected to four Application Specific Integrated Circuits (ASICs), VATA451.2, and the signal is read out from both the cathode and anode side. Each ASIC consists of 64 channels containing a preamplifier, a fast shaper for self-triggering, and a slow shaper for pulse height measurement \citep{watanabe2014si}.
To reduce noise, a floating readout method is adopted \citep{takeda2007development}. 
% The detector is directly connected to the ASICs via DC coupling, with the anode side operating such that the readout circuit floats relative to the bias voltage.
% The detector is directly connected to the ASIC by DC coupling, and the anode side operates on a floating ground at a bias voltage.
The detector is directly connected to the ASIC via DC coupling, and the anode side is referenced to a floating ground set at the high bias voltage.
The digital signal processing section of each ASIC is electrically isolated by a digital isolator and is connected to the FPGA on the DIO board.

\subsection{Data Electronics and Power Board}
\begin{figure}[htb]
  \begin{center}
  \includegraphics[width=1.\hsize]{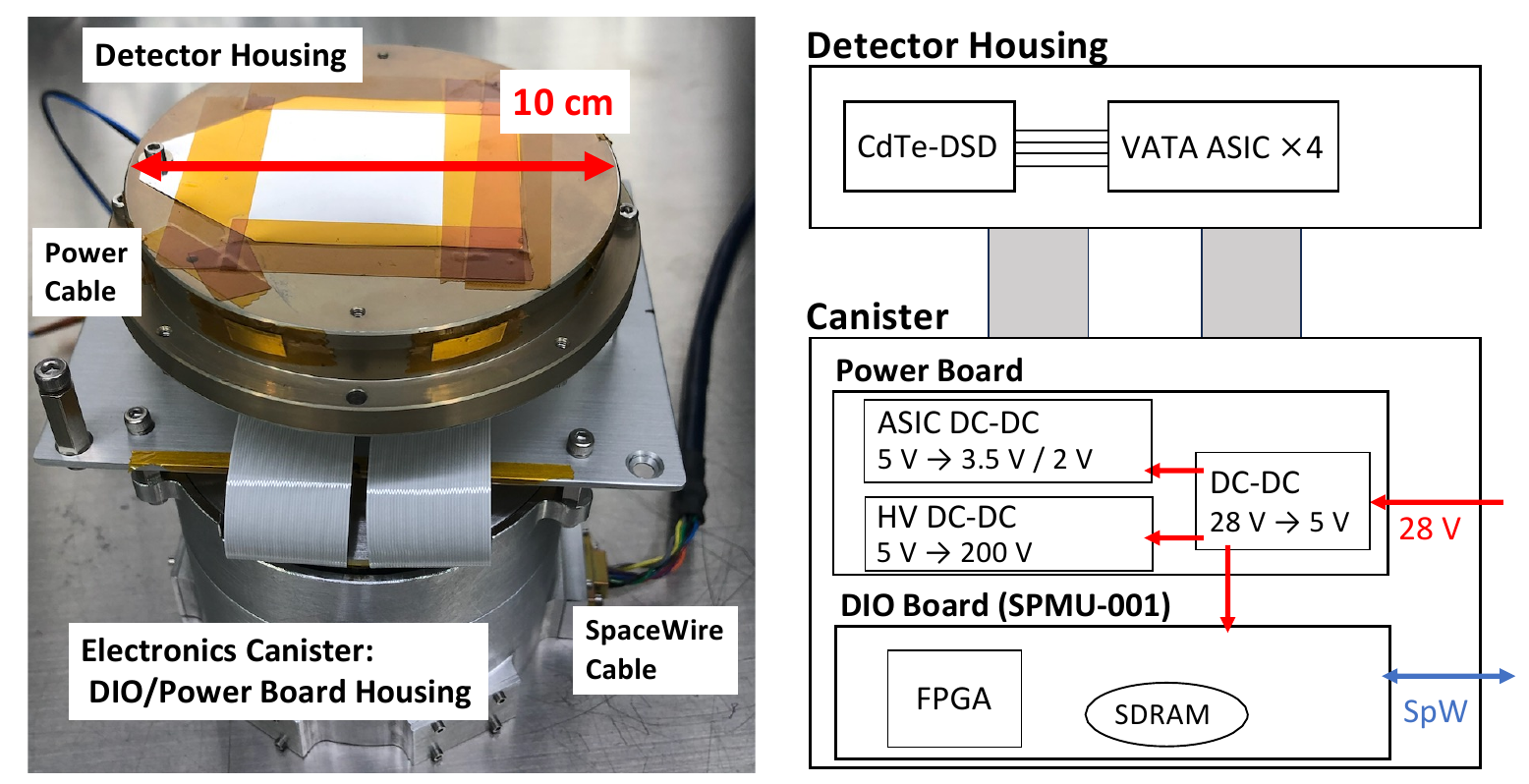}
  \caption{A picture and schematic diagram of the detector and DIO/power-supply board housing (``Electronics Canister"). The power for ASIC, DIO board, and bias voltage is created from the input voltage of 28 V by the power board. The data acquisition of the detector is conducted by the DIO board (SPMU-001), which can be controlled by the SpaceWire command.}
  \label{dsd_housing}
  \end{center}
\end{figure}

Fig.\,\ref{dsd_housing} shows the housing for the CdTe-DSD detector and the DIO/power board. The detector is installed in a circular housing with a diameter of 10 cm, while the DIO/power board is enclosed in a cylindrical container called the ``Electronics Canister". They are connected by two flexible flat cables.
The power board uses a DC-DC converter to generate 5 V from the 28 V input voltage to produce power for the DAQ Board. Additionally, it generates and supplies power for the ASICs and provides the bias voltage for the detector.
The DIO board consists of an FPGA board called SPMU-001, which controls the detector and writes data from the detector to SDRAM. Specific ASIC parameters and measurement modes can be adjusted by rewriting SDRAM values via the SpaceWire protocol \cite{esa_spacewire, spw_userguide}.

The SPMU-001 is a general-purpose FPGA board for physical measurement jointly developed by Shimafuji Electric Incorporated\footnote{\url{http://www.shimafuji.co.jp/en/}} and Kavli IPMU/The University of Tokyo.
In addition to a Spartan-7 FPGA and 128 Mbyte DDR2-SDRAM, high-speed communication interfaces such as SpaceWire and Ethernet are implemented on a Raspberry Pi-sized board.
A SpaceWire-AXI bridge port is included, allowing access to SDRAM via the SpaceWire Remote Memory Access Protocol (RMAP) \cite{esa_rmap}.
The board also features an expansion connector for the Raspberry Pi 4, enabling direct power supply from the SPMU-001 to the Raspberry Pi 4.
Additionally, a SpaceWire-Ethernet bridge port allows the Raspberry Pi 4 to transmit SpaceWire packets via Ethernet, enabling software running on the Raspberry Pi 4 to control the system.
Furthermore, a SpaceWire expansion interface board has been developed, which can be installed on top of the SPMU-001 to support up to six SpaceWire physical ports.

In the experiment setup, the SPMU-001 serves as the DIO board to control the CdTe-DSD within the electronics canister.
Additionally, we developed a compact readout system for controlling and acquiring data from the CdTe-DSD.
This system utilizes the three-board set comprising the SpaceWire interface board, SPMU-001, and a Raspberry Pi 4, with software developed to run on the Raspberry Pi 4 to send SpaceWire commands to the DIO board.
With the new data acquisition system, the detector achieves a total count rate capability of up to 5000 counts/s, satisfying the performance requirements of the FOXSI-4 mission.
The details of the data acquisition system are described in \cite{nagasawa2024doctor, minami2024hard}.

\section{Spectral Performance} \label{sec:spec}
In this section, we evaluate the spectral performance of wide-gap CdTe-DSDs by irradiating them with X-rays from radioisotope (RI) sources. 
The detector is placed in a thermostatic chamber maintained at -20${}^\circ$C, and the bias voltage of 200 V is supplied by the power board.
The X-rays from the RI source irradiate on the cathode side of the detector.

\subsection{Low-Energy Trigger Efficiency} \label{sec:let}
The detection efficiency at low energies is primarily determined by the trigger efficiency of the fast shaper in the ASIC.
The threshold value, denoted as $V_{\mathrm{th}}$ (in arbitrary units), can be set in the VATA451.2 ASIC control register.
However, due to uncertainty in the fast shaper, photons with energy below the nominal discriminator level may trigger, while photons with energy above the nominal level may fail to trigger.
Therefore, evaluating the low-energy trigger efficiency is crucial for accurately assessing the detection efficiency at lower energies.

\begin{figure}[htb]
  \begin{center}
  \includegraphics[width=1.0\hsize]{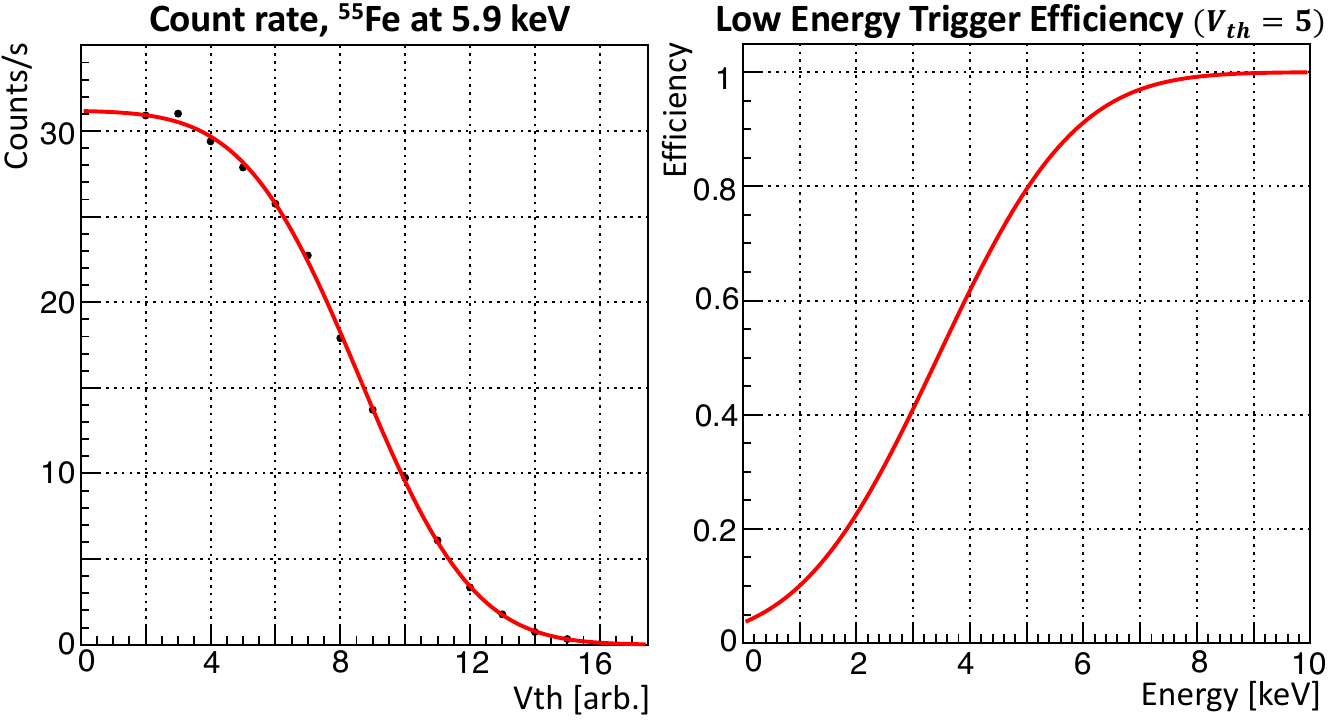}
  \caption{Left: Count rate of 5.9 keV peak of ${}^{55}$Fe with several trigger threshold values $V_{th}$. The black dots represent the measured data, and the red line represents the fitted function of Equation \ref{math_let}. Right: Estimated low energy trigger efficiency for incident photon energy with $V_{\mathrm{th}}=5$.}
  \label{spec_let}
  \end{center}
\end{figure}

To evaluate the low-energy trigger efficiency of the wide-gap CdTe-DSD, we employed the same methodology used for FOXSI-2 CdTe-DSD \cite{ishikawa2016fine}.
We assume that the threshold energy $E_{\mathrm{th}}$ (in keV unit) is linearly related to the ASIC threshold parameter $V_{\mathrm{th}}$ (in arbitrary unit)  by $E_{\mathrm{th}} = G \times V_{\mathrm{th}}$, where $G$ is a constant gain parameter. By varying $V_{\mathrm{th}}$, we collected data using X-ray from a ${}^{55}$Fe RI source and measured the count rate of 5.9 keV line peak (4--8 keV), as shown in Fig.\,\ref{spec_let}.
As we increased $V_{\mathrm{th}}$, the count rate of the 5.9 keV peak decreased, reflecting the reduced trigger efficiency for photons at this energy due to the higher threshold settings.
Assuming that the pulse height spectrum for the fast shaper of the 5.9 keV peak has a Gaussian shape with a resolution of $\sigma_{\mathrm{trig}}$, the count rate $C$ in the 5.9 keV peak can be expressed by \citep{ishikawa2016fine} :
\begin{align}
  C &= C_0 \left[ 1 - \frac{1}{\sqrt{2\pi\sigma_{\text{trig}}}} \int_{-\infty}^{E_{\text{th}}} \exp \left\{ -\frac{(E - 5.9\ \text{keV})^2}{2\sigma_{\text{trig}}^2} \right\} \, dE \right], \\
   &= \frac{C_0}{2} \operatorname{erfc}\left( \frac{GV_{\text{th}} - 5.9\ \text{keV}}{\sqrt{2}\sigma_{\text{trig}}} \right),
   \label{math_let}
\end{align}
where $E$ is an incident photon energy, $C_0$ is the count rate for 100\% trigger efficiency, and erfc($x$) is a complementary error function.
By fitting the experimental relationship between the count rate $C$ and the threshold voltage $V_{\mathrm{th}}$ using Eq.\,\ref{math_let}, we obtained the parameters of $G \sim 0.689$, $\sigma_{\text{trig}} \sim 8.56$ and $C_0 \sim 30.7$.
Finally, the low-energy trigger efficiency $LETE(E)$ at an incident photon energy $E$ can be calculated using the obtained parameters $G$ and $\sigma_{\text{trig}}$ with the following expression:
\begin{equation}
  LETE(E) = \frac{1}{2} \left[ \text{erf}\left( \frac{E - GV_{\text{th}}}{\sqrt{2}\sigma_{\text{trig}}} \right) + 1 \right]
\end{equation}
where erf($x$) is a error function. This function describes the probability that a photon with energy $E$ will trigger the detector, accounting for the resolution of the fast shaper.
The estimated low-energy trigger efficiency with the FOXSI-4 flight setting of $V_{\mathrm{th}} = 5$ is shown in the right panel of Fig.\,\ref{spec_let}. Since the photoelectric absorption efficiency below 20 keV is approximately 100\% for 750 $\mathrm{\mu m}$ thick CdTe, 
we confirmed that the detection efficiency is sufficient in the 4--20 keV range, which is the observation bandwidth of FOXSI-4.

\subsection{Ratio of Charge Sharing Events}\label{sec_ratio_eve}
Table.\,\ref{charge_ratio_Co57} shows the percentage of charge-sharing events for each strip and gap width region of wide-gap CdTe-DSD. We selected events corresponding to the 14 keV peak of ${}^{57}$Co within the energy range of 12-16 keV. 
% The energy threshold of each strip was set to 1.5 keV, the same as in the FOXSI-3 CdTe-DSD \cite{furukawa2020imaging}.
For analysis, the energy threshold for detection on the strips was set to 4 keV, with energies on neighboring strips considered as charge sharing if they exceeded the threshold of 1.5 keV ($\sim 5 \sigma$ of the pedestal width), the same as in the FOXSI-3 CdTe-DSD \cite{furukawa2020imaging}.
\begin{table}[h]
  \centering
  \caption{Percentage of charge-sharing events (${}^{57}$Co 14 keV peak)}
  \label{charge_ratio_Co57}
  \begin{tabular}{l|lll}
  \hline
  Wide-gap CdTe-DSD & Single-strip & Double-strip & $\geq 3$ strip\\ 
   & Cathode / Anode & Cathode / Anode & Cathode / Anode\\ \hline \hline
  (A) 60~$\mathrm{\mu m}$ strip-pitch & 52.98\% / 51.18\%& 46.58\% / 48.80\% & 0.44\% / 0.02\%\\
  (B) 80~$\mathrm{\mu m}$ strip-pitch & 44.83\% / 62.71\%& 54.80\% / 37.27\% & 0.37\% / 0.02\%\\
  (C) 100~$\mathrm{\mu m}$ strip-pitch & 40.68\% / 70.26\%& 58.95\% / 29.73\% & 0.37\% / 0.01\%\\
  \hline
  \end{tabular}
\end{table}

In the FOXSI-3 CdTe-DSD (strip width 50 $\mathrm{\mu m}$ and gap width 10 $\mathrm{\mu m}$), the proportion of double-strip events was limited to approximately 25\% on the cathode side and 50\% on the anode side \citep{furukawa2020imaging}. In contrast, for the wide-gap CdTe-DSD with the same strip-pitch (strip width 30 $\mathrm{\mu m}$ and gap width 30 $\mathrm{\mu m}$), the proportion of double-strip events increased to about 47\% on the cathode side. Furthermore, when the gap width was increased from 30 $\mathrm{\mu m}$ to 70 $\mathrm{\mu m}$ while keeping the strip width fixed at 30 $\mathrm{\mu m}$, the ratio of double-strip events on the cathode side increased further to 59\%. This result is consistent with previous findings from a one-sided wide-gap strip detector \cite{nagasawa2023wide}.
On the other hand, for the anode side, the proportion of double-strip events remained approximately 49\% for the same strip-pitch region (strip width 30 $\mathrm{\mu m}$ and gap width 30 $\mathrm{\mu m}$) and decreased to about 30\% for the widest gap region (gap width 70 $\mathrm{\mu m}$). Despite the decrease, the ratio of double-strip events on the anode side remained sufficient for effective position reconstruction.

The proportion of events involving three or more strips is less than 0.5\% within the observation energy range of FOXSI-4 ($< 20~$keV). Therefore, we ignore these events in the subsequent analyses.

\subsection{Averaged Energy Spectra and Charge-loss by Wide-gaps} \label{sec:charge_loss}
To evaluate the spectroscopic performance, we uniformly irradiated the detector on the cathode side with X-rays from the ${}^{241}$Am source.
Whereas pixel-type detectors read signals from only one side, the CdTe-DSD measures induced charge independently from both the cathode and anode sides. We define these measurements as $E_{Cath.}$ and $E_{Anod.}$, respectively.

\begin{figure}[htb]
  \begin{center}
  \includegraphics[width=1.0\hsize]{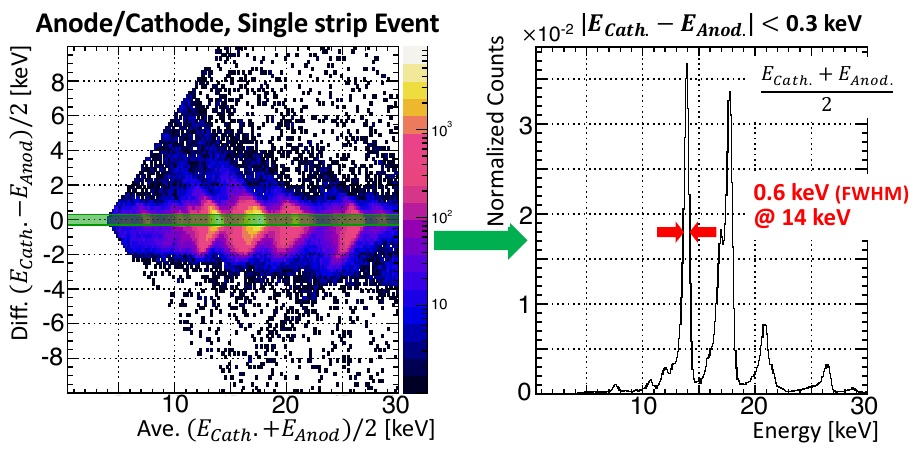}
  \caption{(Left):Relationship between cathode and anode side of energies for single-strip events. (Right): Averaged spectrum for single-strip events in which the energy difference between the anode and cathode side is within 0.3 keV ($\sim 1 \sigma$ of the pedestal).}
  \label{spec_good}
  \end{center}
  \end{figure}

Assuming that the energy resolution is dominated by the noise of the readout electronics on each side, the energy resolution can improve when averaging the energies measured by the cathode and anode side.
Fig.\,\ref{spec_good} shows the averaged energy spectrum ($E = (E_{Cath.}+E_{Anod.})/2$) for single-strip events in which the energy difference between the anode and cathode side is within 0.3 keV ($\sim 1 \sigma$ of the pedestal width).
In this simple subset of events, we achieved a superior energy resolution of 0.6 keV (FWHM) at the 14 keV peak.
However, these events constitute only about 7\% of all detected events.
The overall energy resolution is degraded due to charge loss associated with the wider gaps between strips and the DoI effect \cite{furukawa2020imaging, nagasawa2023wide}.

\begin{figure}[htb]
  \begin{center}
  \includegraphics[width=1\hsize]{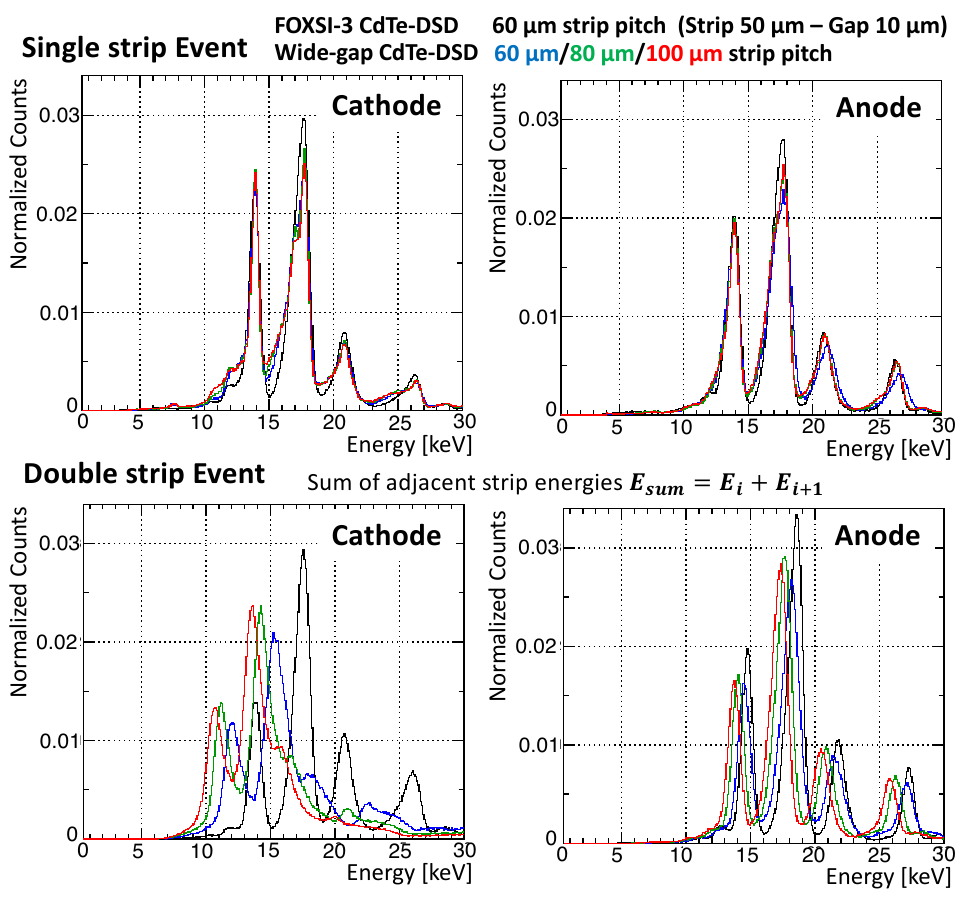}
  \caption{Spectra of ${}^{241}$Am for (Upper) single-strip events and (Lower) double-strip events in each strip/gap region of the wide-gap CdTe-DSD: 60 $\mathrm{\mu m}$, 80 $\mathrm{\mu m}$, and 100 $\mathrm{\mu m}$ strip-pitch regions are shown in red, green, and blue, respectively. The energy spectrum of the FOXSI-3 CdTe-DSD is also overlaid in black for comparison. For double-strip events, the sum of the strip energies is adopted as the photon energy (i.e., $E_{sum} = E_{i} + E_{i+1}$, where $E_{i}$ and $E_{i+1}$ are the detected energies on the $i$-th and $(i+1)$-th strips, respectively).}
  \label{spec_org}
  \end{center}
\end{figure}

Fig.\,\ref{spec_org} shows the energy spectra of the ${}^{241}$Am source obtained for each strip/gap width region in the wide-gap CdTe-DSD.
In these spectra, the sum of the strip energies of adjacent strips $E_{sum}$ is used as a reconstructed energy for double-strip events.
Specifically, we define $E_{i}$ and $E_{i+1}$ as the energies measured from the signals detected in the $i$-th and $(i + 1)$-th adjacent strips, respectively, so that $E_{sum} = E_{i} + E_{i+1}$. 
The energy resolutions (FWHM) at 14 keV peak of ${}^{241}$Am are summarized in Table \ref{simple_fwhm}.

\begin{table}[htb]
  \centering
  \caption{Energy resolutions (FWHM) for single-strip and double-strip events on each side}
  \label{simple_fwhm}
  \small
  \begin{tabular}{l|ll}
  \hline
   & Single-strip& Double-strip \\
   & &$E_{sum}=E_{i}+E_{i+1}$  \\
   & Cathode / Anode & Cathode / Anode \\ \hline \hline
  Wide-gap CdTe-DSD & & \\
  (A) 60~$\mathrm{\mu m}$ strip-pitch & 0.8 keV/1.2 keV & 2.4 keV/1.2 keV\\
  (B) 80~$\mathrm{\mu m}$ strip-pitch & 0.7 keV/1.0 keV & 1.7 keV/1.1 keV\\
  (C) 100~$\mathrm{\mu m}$ strip-pitch & 0.7 keV/1.0 keV & 1.8 keV/1.2 keV\\ \hline
  FOXSI-3 CdTe-DSD & 1.0 keV/1.2 keV & 1.1 keV/1.0 keV\\
  \hline
  \end{tabular}
\end{table}

The obtained spectra for single-strip events indicate that the spectral performance on either the cathode or anode sides in the wide-gap CdTe-DSD remains sufficient for achieving an energy resolution below 1 keV, even in the widest-gap regions, except on the anode side with a 60 $\mathrm{\mu m}$ strip pitch.
The energy resolution is improved compared to the FOXSI-3 CdTe-DSD due to the upgraded ASICs and reduced strip electrode size, which lowers the detector capacitance. However, a low-energy tail component becomes increasingly noticeable in the spectra for the widest-gap regions.

In contrast, the spectral performance for double-strip
events deteriorated, with the peaks shifting to lower energies. This effect is more significant on the cathode side and in the wider gap regions. 
To elucidate the cause of the observed deterioration, we examined the energy relationships between adjacent strips for double-strip events, as shown in Fig.\,\ref{spec_rel_foxsi3}.
For the FOXSI-3 CdTe-DSD, there is a linear relationship between the energies detected in adjacent strips ($E_{sum} = E_{i} + E_{i+1}$ = const.). This implies that a simple summation of the energies on adjacent strips is a reasonable method to reconstruct incident X-ray energies. 
In contrast, for the wide-gap CdTe-DSD, the sum of the energies deviates from the constant and becomes smaller, especially on the cathode side and in the widest gap regions. This trend is also evident for events where the energies of adjacent strips are comparable ($E_{i} \simeq E_{i+1}$). 
Such events occur when photons interact with the detector near the center of the gap.
Thus, simply summing the adjacent strip energies is not appropriate for energy reconstruction in this case.
This charge loss in the gap region has also been reported in previous studies on one-sided wide-gap strip detectors \cite{nagasawa2023wide} and Cadmium-Zinc-Telluride (CdZnTe) pixel detectors \cite{koch2021charge, abbene2018digital}.

Furthermore, the DoI dependence also affects the energy measurement.
When photons interact deeper within the detector, the total measured energy on the cathode electrodes decreases, creating a low-energy tail in the spectrum \cite{furukawa2020imaging, minami20232mmthick}.
Therefore, it is necessary to develop a new energy reconstruction method to compensate for the charge loss caused by wide gaps and DoI effects, thereby enhancing detection efficiency while maintaining spectral performance.

% \clearpage

\subsection{Energy Reconstruction for Charge-loss using Adjacent Strip Energies} \label{sec:methode_chargeloss}
\begin{figure}[htb]
  \begin{center}
  \includegraphics[width=.82\hsize]{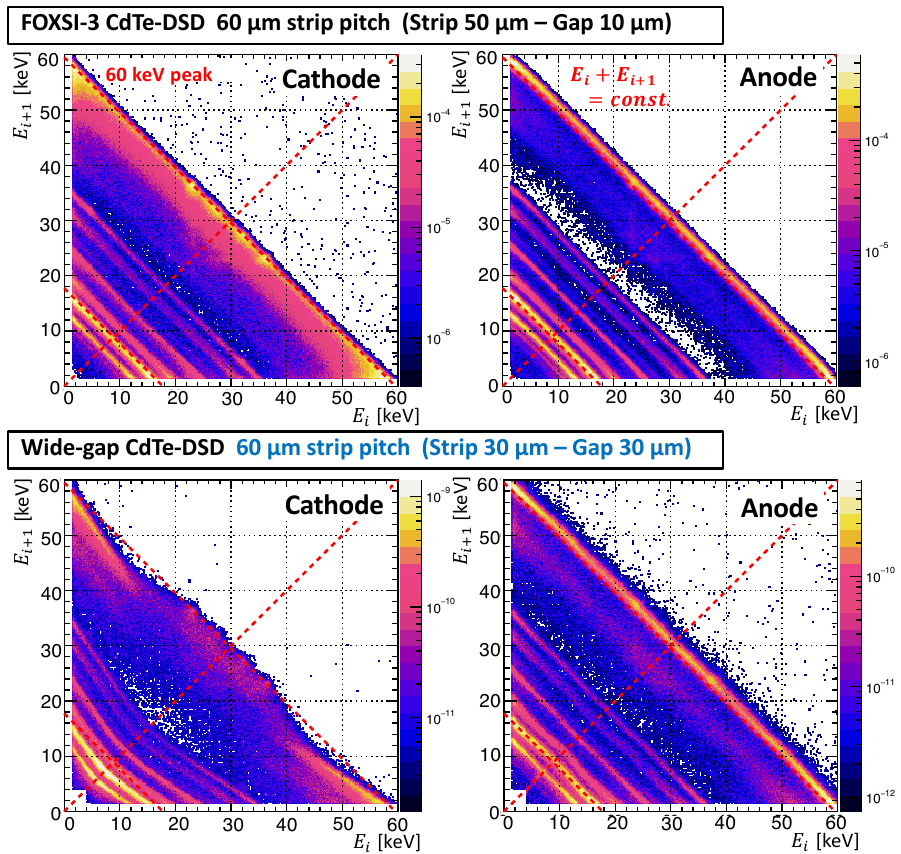}
  \caption{Relationships between the energies detected on adjacent strips for (Upper) the FOXSI-3 CdTe-DSD and (Lower) the wide-gap CdTe-DSD for double-strip events of ${}^{241}$Am. The four points in the 60 keV peak line are the X-ray fluorescence peaks of Cd and Te at 23 keV and 27 keV.}
  \label{spec_rel_foxsi3}
  \end{center}
% \end{figure}

% \begin{figure}[htb]
  \begin{center}
  \includegraphics[width=0.82\hsize]{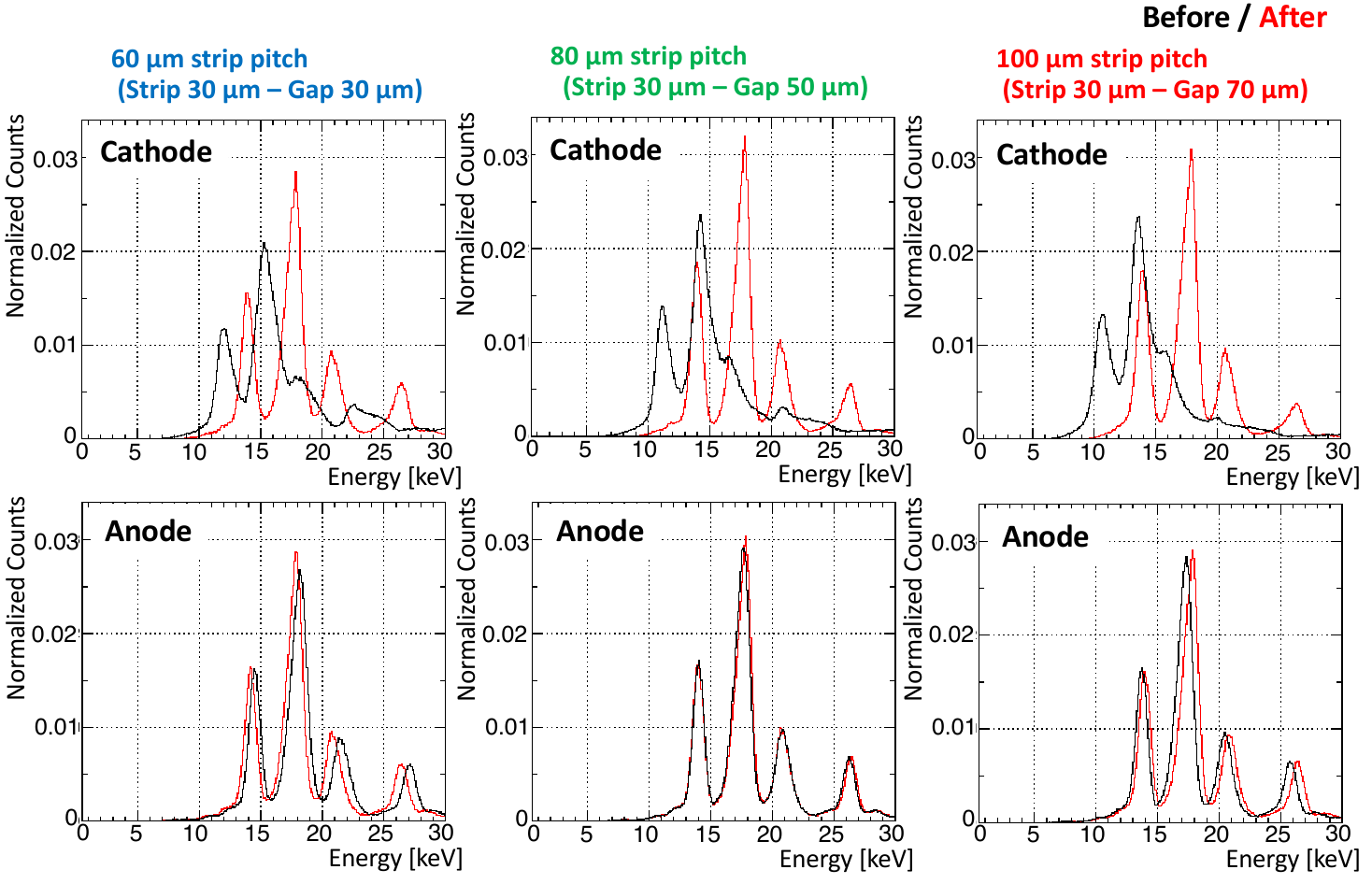}
  \caption{${}^{241}$Am spectra of the (Black) sum of the strip energies $E_{sum}$ and the (Red) reconstructed energy $E_{mod}$ for double-strip events in each strip/gap width region of the wide-gap CdTe-DSDs. All double-strip events are included, demonstrating the improvement after applying the energy reconstruction method.}
  \label{spec_rel_afterspect}
  \end{center}
\end{figure}

The spectra in Fig.\,\ref{spec_org} and the relationships shown in Fig.\,\ref{spec_rel_foxsi3} suggest that part of the charge induced by an incoming photon is lost during charge collection in the electrodes, especially on the cathode side. Moreover, the amount of charge loss likely depends on the photon interaction position inside the detector.
Although the photon interaction position cannot be directly obtained from the data, the energy difference between adjacent strips serves as a good alternative parameter for evaluating the amount of charge loss.
Therefore, we can reconstruct the charge loss caused by the wide gaps for double-strip event by empirically determining the relationships between the sum ($E_{sum} = E_{i} + E_{i+1}$) and difference ($E_{diff} = E_{i}-E_{i+1}$) of the adjacent strip energies for double-strip events. 
We utilized the energy reconstruction method for double-strip events, which was developed using a single-sided wide-gap detector \cite{nagasawa2023wide}. Based on the relationship between $E_{sum}$ and $E_{diff}$, we performed a gap loss correction for each strip-pitch region on both the cathode and anode sides. The details of the method are described in \cite{nagasawa2023wide}.

Fig.\,\ref{spec_rel_afterspect} shows the spectra of the sum of the adjacent strip energies $E_{sum}$ and the reconstructed energy $E_{mod}$ for double-strip events of each strip/gap width region. The energy resolutions at 13.9 keV peak of ${}^{241}$Am after charge-loss reconstruction are summarized in Table \ref{charge_fwhm}.
The deterioration of the spectroscopic performance due to charge loss is effectively reconstructed, and the energy resolution improves from 2.4 keV to 1.3 keV (FWHM) at the 13.9 keV peak after reconstruction for the 60 $\mathrm{\mu m}$ strip-pitch region on the cathode side. The energy resolution of the anode side spectra itself does not change, but the gain shift is properly corrected. 

\begingroup
\begin{table}[htb]
  \centering
  \caption{Energy resolution (FWHM) of double-strip events on each side before and after reconstruction for charge-loss at 13.9 keV peak of ${}^{241}$Am}
  \label{charge_fwhm}
  \small
  \begin{tabular}{l|ll}
  \hline
   & Before  & After \\
   & $E_{sum}=E_{i}+E_{i+1}$ & $E = E_{mod}$ \\
   Double-strip Events & Cathode / Anode & Cathode / Anode \\ \hline \hline
  (A) 60~$\mathrm{\mu m}$ strip-pitch & 2.4 keV/1.2 keV & 1.3 keV/1.2 keV\\
  (B) 80~$\mathrm{\mu m}$ strip-pitch & 1.7 keV/1.1 keV & 1.1 keV/1.1 keV\\
  (C) 100~$\mathrm{\mu m}$ strip-pitch & 1.8 keV/1.2 keV & 1.2 keV/1.2 keV\\
  \hline
  \end{tabular}
\end{table}
\endgroup

\subsection{Energy Reconstruction Using Both Detector Sides Energy Measurements for Charge Loss and Depth-of-Interaction Effects} \label{sec:recont_doi}
By utilizing adjacent strip energy information, we successfully reconstructed the charge loss for double-strip events.
In this section, we extend our method by utilizing energies from both the cathode and anode sides to reconstruct the charge loss for single-strip events and correct for the DoI effect, aiming to improve the energy resolution further.

\begin{figure}[htb]
  \begin{center}
  \includegraphics[width=0.85\hsize]{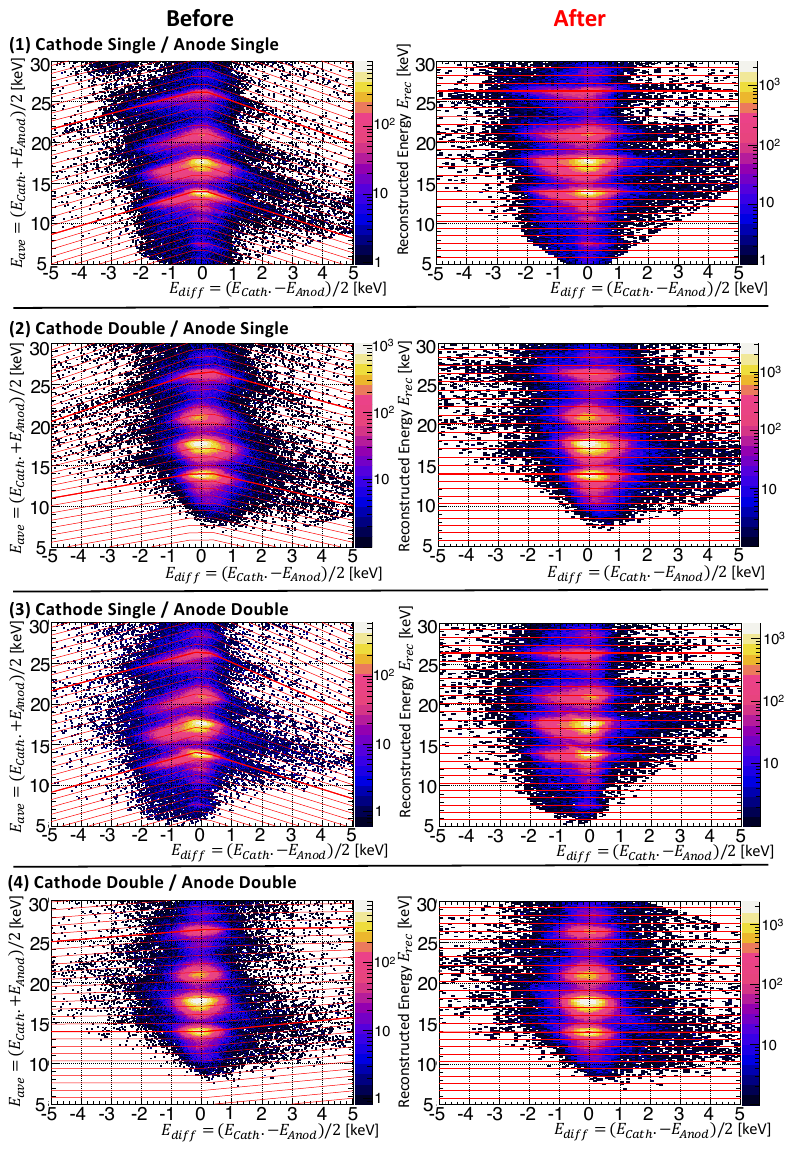}
  \caption{(Left): Relationships between the energy difference $E_{diff} = (E_{Cath.} - E_{Anod.})/2$ and the average energy $E_{ave} = (E_{Cath.} + E_{Anod.})/2$ from both cathode and anode sides, divided by each charge-sharing case. The red lines represent the linear fits used for gain correction. (Right): After applying the gain correction factors derived from the linear functions, the reconstructed energy becomes constant with respect to $E_{diff}$.}
  \label{spec_doi_corr}
  \end{center}
\end{figure}

Fig.\,\ref{spec_doi_corr} shows the relationship between the average energy ($E_{ave} = (E_{Cath.}+E_{Anod.})/2$) and the difference in energies ($E_{diff} = (E_{Cath.}-E_{Anod.})/2$) from both the cathode and anode sides for ${}^{241}$Am. 
Since the charge loss for double-strip events has already been corrected using the method in Section \ref{sec:methode_chargeloss},  we construct the reconstruction function by considering four different charge-sharing cases (single-strip/double-strip) on both the cathode and anode sides.
As shown in Fig.\,\ref{spec_doi_corr}, there are two tail components extending to the left ($E_{Cath.} < E_{Anod.}$) and right ($E_{Cath.} > E_{Anod.}$) sides.

The tail structure on the left side indicates that the energy on the cathode side is smaller than that on the anode side, which is especially noticeable for the cathode-side single-strip events (cases 1 and 3).
For the cathode-side single-strip and the anode-side double-strip events case (3), the charge loss due to the wide gap on the anode side has already been reconstructed using the method described in Section \ref{sec:methode_chargeloss}. 
However, the gain correction for the charge loss has yet to be applied on the cathode side.
If an incident photon interacts near the center of the strip on the cathode side, the charge loss is minimal, resulting in $E_{diff}\sim 0$. However, if the interaction occurs near the center of the gap, significant charge loss occurs on the cathode side, leading to $E_{diff} < 0$ and contributing to the left-side tail structure.
Additionally, the DoI effect for higher-energy photons ($E \gtrsim 20$ keV) causes a further reduction in the cathode-side energy due to hole trapping, enhancing the left-side tail.

Conversely, the right-side tail indicates that the energy on the anode side is smaller than that on the cathode side, particularly for anode-side single-strip events (cases 1 and 2).
The same analogy can explain the cause of this tail structure for case (2): if a photon interacts near the center of a strip on the anode side, the charge loss is minimal, and $E_{diff}\sim 0$. However, interactions near the center of a gap result in a significant charge loss on the anode side, leading to $E_{diff} > 0$.
Furthermore, for single-strip events on both sides (case 1), the energy difference depends on whether the interaction point is near a gap or a strip on each side, extending both the left and right tail structures.

To correct these tail structures, we employed a similar charge loss correction approach as in Section \ref{sec:methode_chargeloss}. Using the energy difference of both sides $E_{diff}$ as a parameter, we fitted the peak position of the left and right tail structures for each line peak of ${}^{241}$Am with a linear function ($a_0 + a_1E_{diff}$), as indicated by the red lines of Fig.\,\ref{spec_doi_corr} (left). 
We determined the linear functions separately for three regions: (1) the left side ($E_{diff}< E_{low}$), (2) the right side ($E_{diff}>E_{high}$), and (3) the center ($E_{low} \leq E_{diff} \leq E_{high}$). 
The threshold energies $E_{low}$ and $E_{high}$ were also fitted for each charge-sharing event case.
We calculated the gain correction factors based on the best-fit linear functions for each line peak of ${}^{241}$Am. This ensures that the reconstructed energy becomes constant with respect to $E_{diff}$, as shown in Fig.\,\ref{spec_doi_corr} (right).
For energies between the line peaks, we determined the gain correction factors using linear interpolation. For energies above the highest peak (26.3 keV) and below the lowest peak (13.9 keV), we applied the same gain correction factors as extrapolations.

\begin{figure}[htb]
  \begin{center}
  \includegraphics[width=0.9\hsize]{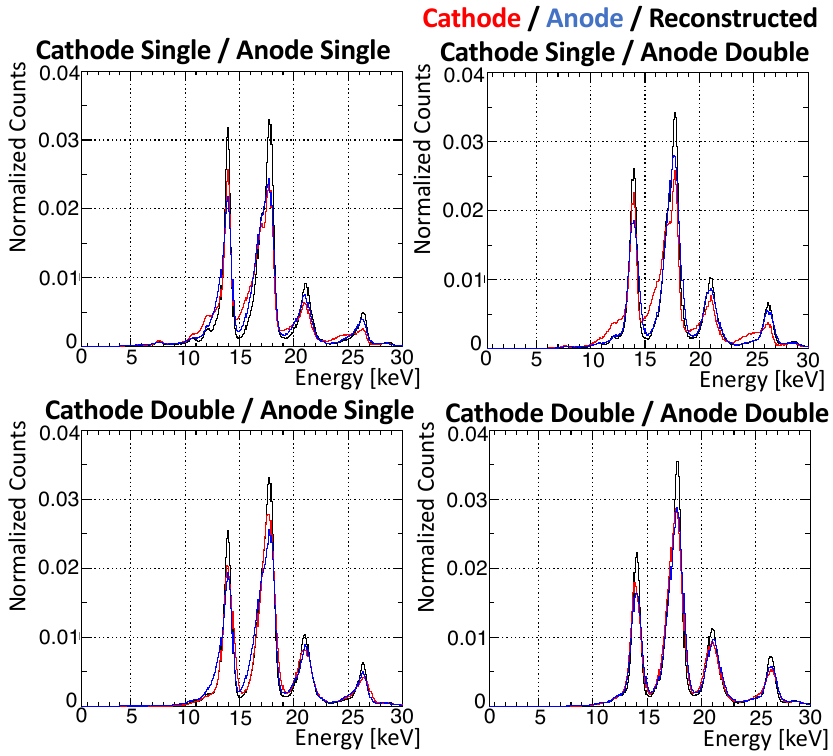}
  \caption{Spectra for each charge-sharing case showing the (Red) cathode-side energy, (Blue) anode-side energy, and (Black) reconstructed energy. The charge-loss correction in Section \ref{sec:methode_chargeloss} is applied for double-strip events. The low-energy tail structures for single-strip events are effectively corrected.}
  \label{spec_doi_afterspect}
  \end{center}
% \end{figure}

% \begin{figure}[htb]
  \begin{center}
  \includegraphics[width=0.9\hsize]{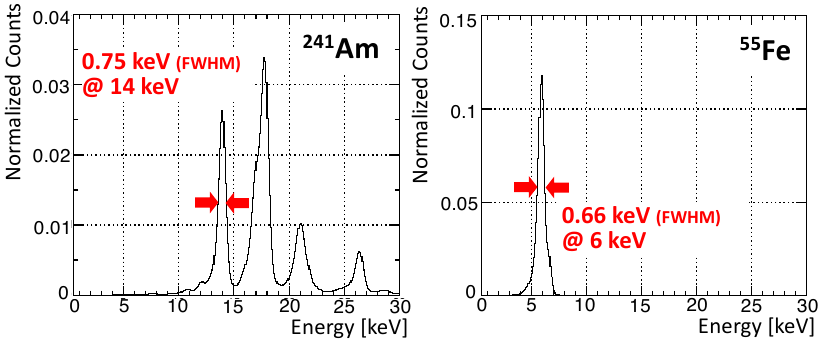}
  \caption{Reconstructed energy spectra of ${}^{241}\mathrm{Am}$ and ${}^{55}\mathrm{Fe}$ using all single-strip and double-strip events. The spectra from all channels are combined after applying charge-loss and DoI corrections.}
  \label{spec_fin_reconst}
  \end{center}
\end{figure}

Fig.\,\ref{spec_doi_afterspect} shows the reconstructed spectra for each charge-sharing event case. The energy resolutions at 13.9 keV peak of ${}^{241}$Am for the reconstructed spectra are summarized in Table \ref{doi_fwhm}.
The low-energy tail structures for single-strip events are effectively corrected, and the energy resolutions are also improved. 
Fig.\,\ref{spec_fin_reconst} shows the reconstructed spectra of ${}^{241}$Am and ${}^{55}$Fe using all singe-strip and double-strip events. 
To reconstruct ${}^{55}$Fe spectra, the same correction functions derived from the best-fit measurements with the ${}^{241}$Am is used. 
The energy resolutions (FWHM) of 0.75 keV at the peak of 13.9 keV and 0.66 keV at the peak of 5.9 keV are obtained.
Therefore, we confirm that sufficient spectroscopic performance is achieved using all single-strip and double-strip events by applying the charge-loss and DoI corrections, despite the wider gaps compared to the FOXSI-3 CdTe-DSDs.

\begin{table}[htb]
  \centering
  \caption{Energy resolution (FWHM) of reconstructed spectra using both detector sides energy measurements at 13.9 keV peak of ${}^{241}$Am}
  \label{doi_fwhm}
  \small
  \begin{tabular}{lll|lll}
  \hline
  \multicolumn{3}{l|}{Event Case} & Cathode & Anode & Reconstructed  \\ 
  &Cathode & Anode & & &\\ \hline \hline
  (1) &Single & Single & 0.7 keV  & 1.0 keV & 0.6 keV \\
  (2) &Double & Single & 1.2 keV  & 1.2 keV & 0.8 keV \\
  (3) &Single & Double & 0.7 keV  & 1.1 keV & 0.7 keV \\
  (4) &Double & Double & 1.0 keV  & 1.1 keV & 0.8 keV \\
  \hline
  \end{tabular}
\end{table}

% \clearpage

\subsection{Uniformity of Spectral Performance} \label{sec:uniform}
\begin{figure}[htb]
  \begin{center}
  \includegraphics[width=.72\hsize]{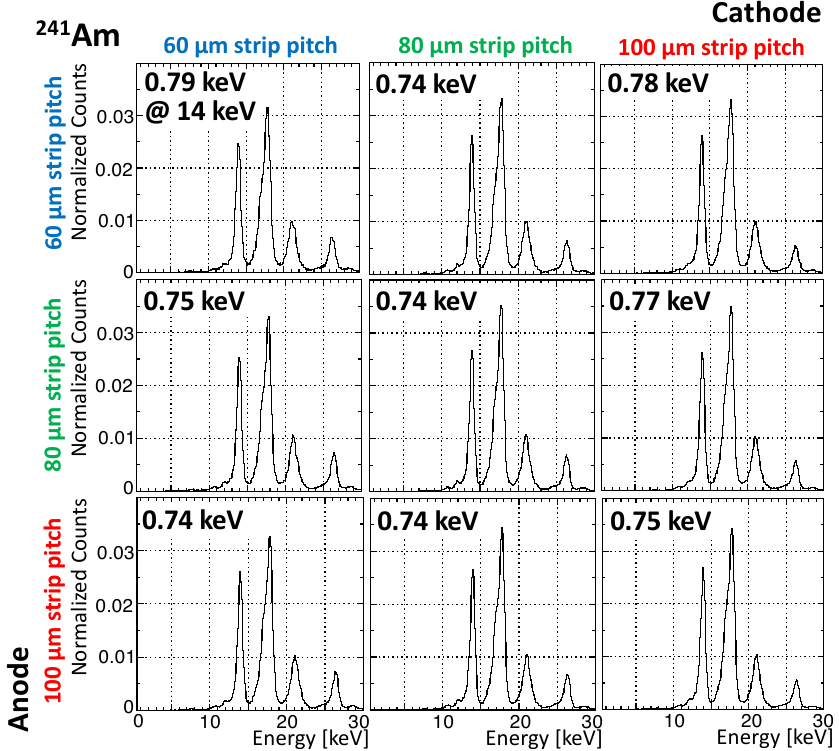}
  \caption{Reconstructed spectra of ${}^{241}$Am for each strip/gap width region. The energy resolution (FWHM) at the 13.9 keV peak is shown above each panel.}
  \label{spec_fin_am241}
  \end{center}
%   \end{figure}
% \begin{figure}[htb]
  \begin{center}
  \includegraphics[width=.72\hsize]{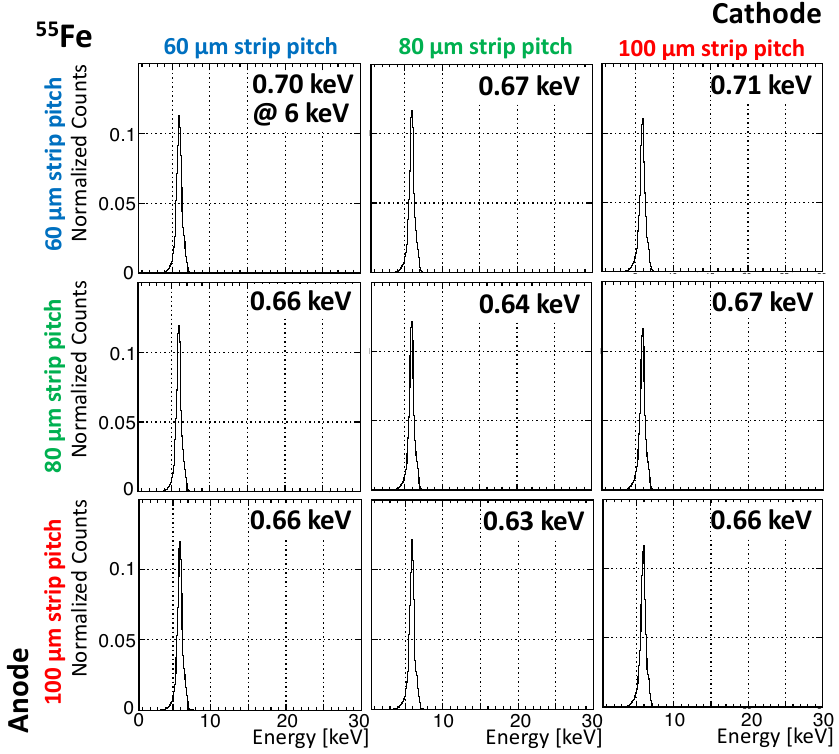}
  \caption{Reconstructed spectra of ${}^{55}$Fe for each strip/gap width region. The energy resolution (FWHM) at the 5.9 keV peak is shown above each panel.}
  \label{spec_fin_fe55}
  \end{center}
\end{figure}

In this section, we examined the uniformity of the spectral performance of the detector across different strip and gap width regions. Fig.\,\ref{spec_fin_am241} and \ref{spec_fin_fe55} show the final reconstructed spectra for each strip/gap width region of ${}^{241}$Am and of ${}^{55}$Fe, respectively, after applying the reconstruction method using both sides energy measurements as described in Section \ref{sec:recont_doi}. The energy resolution is stable across all strip/gap width regions, ranging from 0.74 keV to 0.79 keV at the 13.9 keV peak of ${}^{241}$Am and from 0.63 keV to 0.71 keV at 5.9 keV peak of ${}^{55}$Fe.
Therefore, we confirm that the wide-gap CdTe-DSD exhibits consistent and sufficient spectroscopic performance across all strip and gap width regions.

%\clearpage

\section{Characterization of Charge Sharing for Sub-strip Position Reconstruction} \label{sec:spring8}
In this section, we conducted a synchrotron beam scanning test at the Spring-8 Synchrotron Radiation Facility to evaluate the charge-sharing properties of the wide-gap CdTe-DSD.
Based on this result, we develop a sub-strip-level position reconstruction method by utilizing the sharing energy information between adjacent strips.

\subsection{Experimental Setup for Beam Scanning Test at SPring-8}
We used the synchrotron X-ray beamline BL20B2 at the SPring-8 Synchrotron Radiation Facility 
\citep{goto2001construction} in Japan.
The BL20B2 beamline has a total length of 215 m from the bending-magnet source to the end in the experimental hutch, providing a parallel incident X-ray beam.
A monochromatic X-ray beam is extracted using the SPring-8 standard double-crystal monochromator \cite{yabashi1999spring} located at 36.8 m from the source.
The X-ray energy range is adjustable from 4.4 to 113 keV, with an energy uncertainty of $\Delta E/E \sim 10^{-4}$, which is much smaller than the energy resolution of the detector.

\begin{figure}[htb]
  \begin{center}
  \includegraphics[width=1.0\hsize]{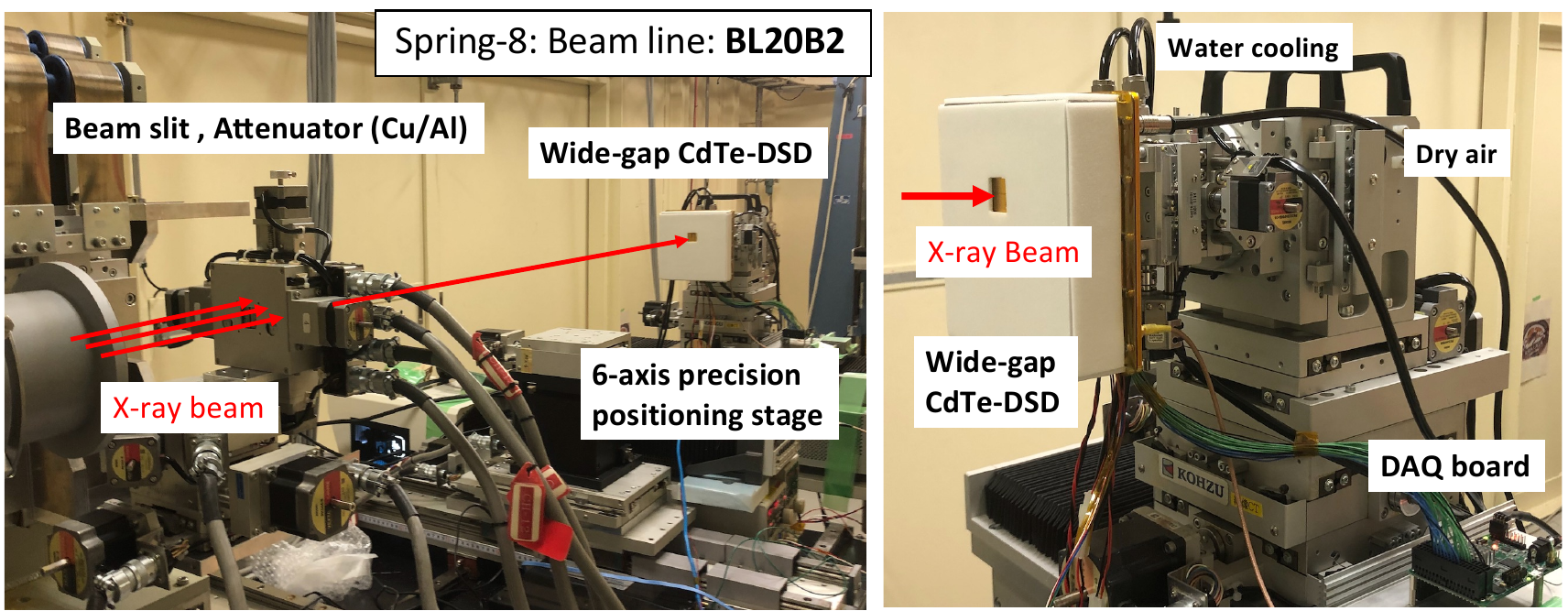}
  \includegraphics[width=1.0\hsize]{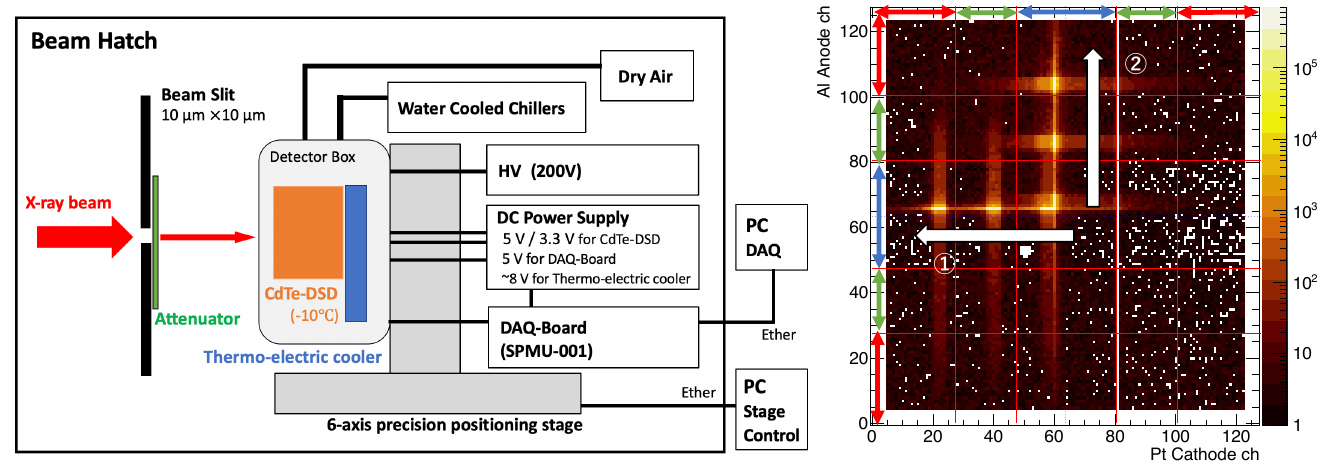}
    \caption{(Top) Pictures and (Lower left) schematic diagram of the experimental setup for the beam scanning test. (Lower right) The detector image of all scanning data. 
    The white arrow represents the beam scanning direction, and the colored arrows on the sides represent the different strip-pitch regions.
    Channel 81 on the cathode side is not connected in this prototype CdTe-DSD used for the beam scanning test.}
    \label{beam_setup_fig}
    \end{center}
\end{figure}

Fig.\,\ref{beam_setup_fig} shows the experimental setup for the beam scanning test.
The monochromatic X-ray beam is collimated to a $10\times10~\mathrm{\mu m^2}$ square using a four-quadrant tantalum slit.
The beam intensity is attenuated by using Cu and Al plates.
The CdTe-DSD is mounted on a 6-axis precision positioning manipulator (MPS-SSSD-B010, KOHZU Precision Co., Ltd.), which can be controlled in 0.5 $\mathrm{\mu m}$ step with a resolution of 0.1 $\mathrm{\mu m}$.
The detector is cooled to $-10{}^{\circ}$C by a thermo-electric cooler, with the cooler side attached to the detector and the hotter side cooled by a water-cooled chiller.
Dry air flows inside the detector box to prevent condensation, and a bias voltage of 200 V is applied.

\begin{figure}[htb]
  \begin{center}
  \includegraphics[width=1\hsize]{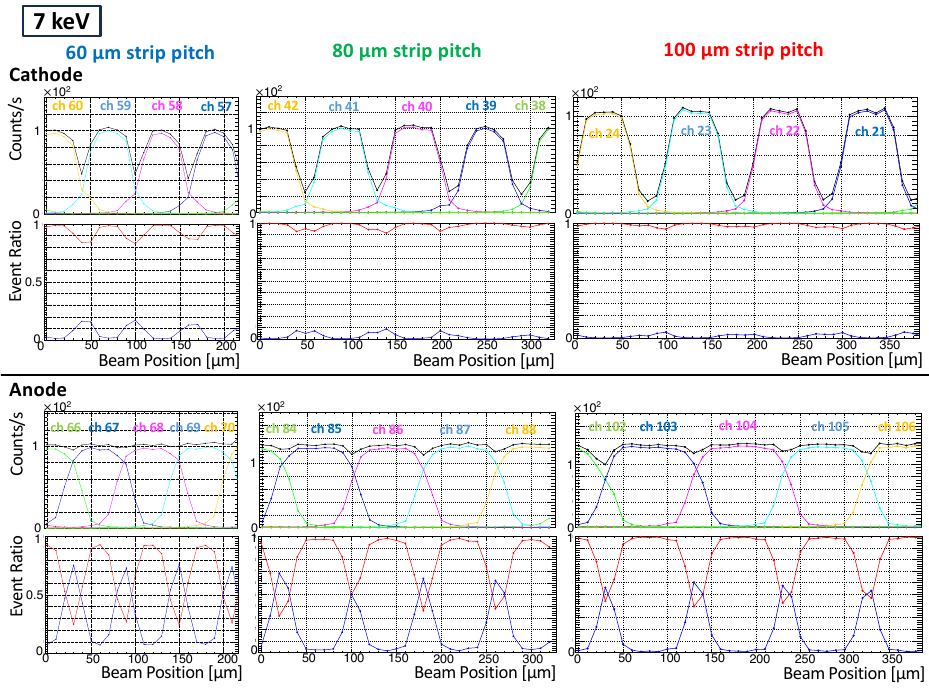}
  \caption{Beam scanning results for the 7 keV beam on the cathode and anode sides for each strip/gap width region. Top: Total count rates (Black) and individual count rates by channel (Colored). For double-strip events, one count is added to both channels. Bottom: Ratio of single-strip (Red) and double-strip (Blue) events relative to the beam position.}
  \label{beam_cnt_ratio_7keV}
  \end{center}
\end{figure}

\begin{figure}[htb]
  \begin{center}
  \includegraphics[width=1\hsize]{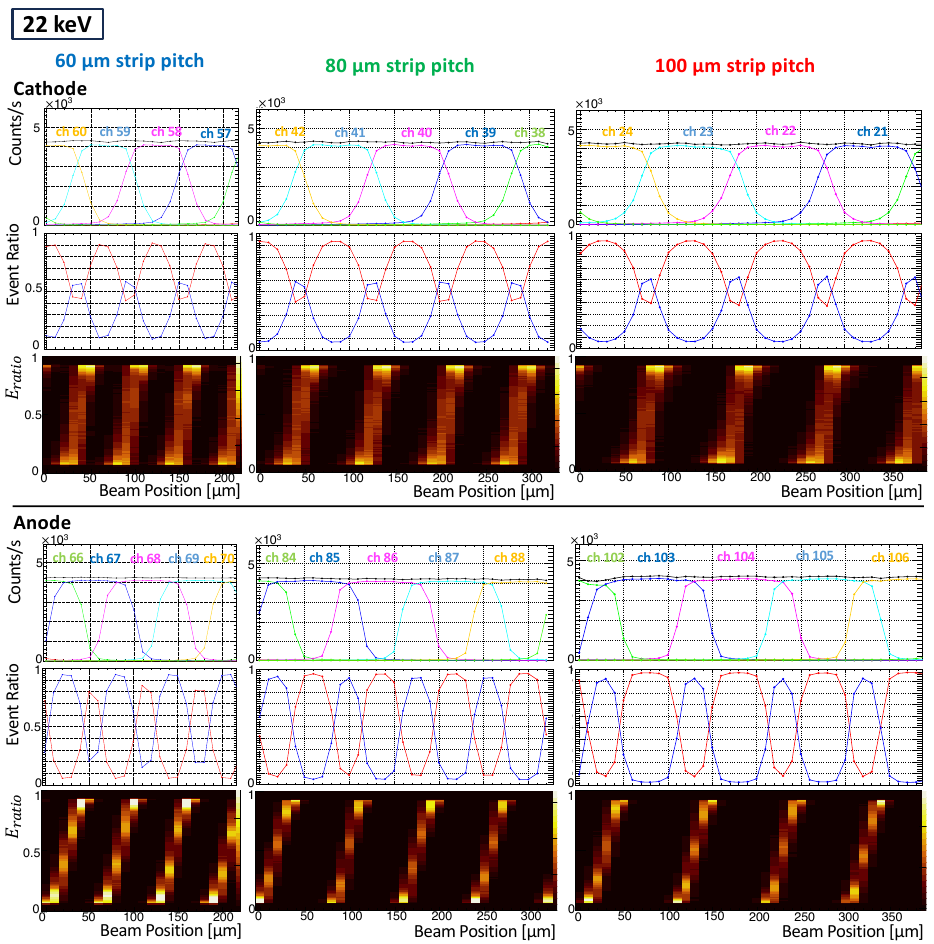}
  \caption{Beam scanning results for the 22 keV beam on the cathode and anode sides for each strip/gap width region. Top: Total count rates (Black) and individual count rates by channel (Colored). For double-strip events, one count is added to both channels. Middle: Ratio of single-strip (Red) and double-strip (Blue) events relative to the beam position. Bottom: Relationship between the detected energy ratio in adjacent strips $E_{ratio} = E_{i+1}/(E_{i} + E_{i+1})$ and the incident X-ray position $X_{pos}$.}
  \label{beam_cnt_ratio_22keV}
  \end{center}
\end{figure}

The purpose of this experiment is to examine how the X-ray beam position affects the detector response across different strip and gap width regions. Specifically, we aim to evaluate the uniformity of detection efficiency, the ratio of charge-sharing events, and the variations in sharing energy between adjacent strips.
To achieve this, we conducted scans perpendicular to four strips in each strip-pitch region on both the cathode and anode sides, forming an L-shape pattern, as shown in Fig.\,\ref{beam_setup_fig} (lower right).
Alignment was adjusted to ensure scanning along the center of the strip on the opposite side.
Beam energies were set to 7 keV, 14 keV, and 22 keV. The step size was 10 $\mathrm{\mu m}$, with integration times of 55 seconds for the 14 keV and 22 keV beam and 120 seconds for the 7 keV beam due to the lower beam intensity.
The results of each strip and the gap width region on both the cathode and anode sides are summarized in Fig.\,\ref{beam_cnt_ratio_7keV}, Fig.\,\ref{beam_cnt_ratio_14keV} and Fig.\,\ref{beam_cnt_ratio_22keV} for beam energies of 7 keV, 14 keV, and 22 keV, respectively, providing insight into the uniformity of detection efficiency and charge-sharing behavior.

\subsection{Results} 
\subsubsection{Uniformity of Detection Efficiency} \label{sec:beam}
To examine the uniformity of detection efficiency, we analyzed the count rate variations in different regions of strip and gap width.
The top panels of Fig.\,\ref{beam_cnt_ratio_7keV}, Fig.\,\ref{beam_cnt_ratio_14keV} and Fig.\,\ref{beam_cnt_ratio_22keV} show the detected count rates as a function of incident X-ray beam position.
Table \ref{deteff_cathode} and \ref{deteff_anode} summarize the average count rates, standard deviations, and statistical errors for both the cathode and anode sides. The standard deviations and statistical errors are normalized by the average count rates and expressed as percentages.

\begin{table}[h]
  \centering
  \caption{Average count rate and uniformity of detection efficiency on the cathode side}
  \label{deteff_cathode}
  \small
  \begin{tabular}{l|l|l|l}
  \hline
   Cathode & 7 keV & 14 keV & 22 keV\\  \hline \hline
   60~$\mathrm{\mu m}$ strip-pitch
   & 84.7 counts/s & 3822.2 counts/s & 4269.9 counts/s  \\
   ~~std. / stat. err. & 23.2\% / 1.1\% & 2.2\% / 0.3\% & 0.7\% / 0.3\% \\\hline
   80~$\mathrm{\mu m}$ strip-pitch
   & 74.3 counts/s & 3690.2 counts/s & 4266.9 counts/s  \\
   ~~std. / stat. err. & 41.0\% / 1.2\% & 3.1\% / 0.3\% & 0.7\% / 0.3\% \\\hline
   100~$\mathrm{\mu m}$ strip-pitch
   & 67.3 counts/s & 3810.8 counts/s & 4255.5 counts/s  \\
   ~~std. / stat. err.& 55.0\% / 1.3\% &1.9\% / 0.3\% & 0.8\% / 0.3\% \\\hline
  \end{tabular}

  \centering
  \caption{Average count rate and uniformity of detection efficiency on the anode side}
  \label{deteff_anode}
  \small
  \begin{tabular}{l|l|l|l}
  \hline
   Anode & 7 keV & 14 keV & 22 keV\\  \hline \hline
   60~$\mathrm{\mu m}$ strip-pitch
   & 101.6 counts/s & 3701.2 counts/s & 4210.9 counts/s  \\
   ~~std. / stat. err.& 1.7\% / 1.1\% & 0.3\% / 0.3\% & 0.3\% / 0.3\% \\\hline
   80~$\mathrm{\mu m}$ strip-pitch
   & 126.5 counts/s & 3818.2 counts/s & 4224.7 counts/s  \\
   ~~std. / stat. err.& 3.9\% / 0.9\% & 0.4\% / 0.3\% & 0.5\% / 0.3\% \\\hline
   100~$\mathrm{\mu m}$ strip-pitch
   & 128.7 counts/s & 3830.0 counts/s & 4187.3 counts/s  \\
   ~~std. / stat. err.& 3.1\% / 0.9\% & 0.3\% / 0.3\% & 0.7\% / 0.3\% \\\hline
  \end{tabular}
\end{table}

At 14 keV and 22 keV on the anode side, the variation in detection efficiency is consistent with the statistical error, indicating stable detection efficiency at the 10 $\mathrm{\mu m}$ scale with $\sigma \sim 1$\%.
In contrast, the cathode side shows slightly larger variations with $\sigma \sim 3$\%. This may be due to variations in the trigger level of each strip, as the detector is triggered by the strip on the cathode side.
Especially at 14 keV energy on the cathode side, the count rate decreases to $\sim 92$\% at the center of the strip. This reduction is due to absorption by the cathode electrode, which is consistent with a 10\% absorption calculated based on the electrode thickness. Additionally, there may be variations in the trigger level of each strip, as the detector is triggered by the strip on the cathode side.

At 7 keV on the anode side, no significant difference is observed between the variation in detection efficiency and statistical error with $\sigma \sim 4$\%.
In contrast, the cathode side has large variations compared to the statistical error, with detection efficiency decreasing by $\sim$ 50\% in the gap region for the 60 $\mathrm{\mu m}$ strip-pitch region.
This variation is likely due to the trigger configuration.
With the fast shaper threshold set to $V_{th}=5$, the trigger efficiency (as shown in Fig.\,\ref{spec_let}) decreases in regions where charge sharing divides the energy across strips. For example, if a 7 keV X-ray interacts at the center of a gap and splits into two 3.5 keV strip signals, the triggering probability per strip is $\sim 53$\%. 
The detection efficiency is further reduced by charge loss in the gap region, especially as the gap width increases.
However, the detection efficiency changes periodically in response to each strip and gap position. Thus, it is possible to model the variation in detection efficiency based on the results of this experiment.

\subsubsection{Ratio of Charge Sharing} \label{sec:beam_ratio}
To characterize charge-sharing properties across energy bands and strip/gap width regions, we investigated the relationship between the beam incident position and the ratio of charge-sharing events.
The bottom panel of Fig.\,\ref{beam_cnt_ratio_7keV} and the middle panels of Fig.\,\ref{beam_cnt_ratio_14keV} and Fig.\,\ref{beam_cnt_ratio_22keV} show the ratios of single and double-strip events relative to the incident X-ray beam position.
The ratio of charge-sharing events varies periodically across strip and gap positions, suggesting that the charge-sharing properties are common in each strip and have no individual differences. 
In all energy bands, the anode side consistently shows a higher ratio of double-strip events than the cathode side. This is likely because, at energies below approximately 20 keV, photons interact near the cathode side surface of the detector (X-rays irradiate on the cathode side), allowing the charge cloud to diffuse before reaching the anode.

When comparing the results of each energy scan, the 7 keV scan on the cathode side shows that double-strip events make up fewer than 15\% of events, with nearly all events being single-strip due to the lower trigger sensitivity at this energy level, as discussed in Section \ref{sec:beam}.
In contrast, the anode side shows a double-strip event ratio of approximately 50\% at the gap center.
For the 14 keV scan, the double-strip event ratio on the cathode side increases to around 40\% at the gap center, while the anode side exceeds 90\%, reflecting increased charge-sharing due to higher energy.
For the 22 keV scan, the ratio of double-strip events further increases to about 60\% on the cathode side, and the region where double-strip events are dominant (defined as the ``double-strip region" where the double-strip event is $\gtrsim 20\%$) is also extended.

Comparing the difference of each strip/gap width region, on the cathode side, the width of the double-strip region is slightly extended with increasing gap width, leading to a higher ratio of double-strip events.
In contrast, on the anode side, the double-strip region shows little variation. 
For the 14 keV scan, the width of the double-strip region is 50--60 $\mathrm{\mu m}$ for all strip/gap width regions, and the single-strip region is extended as the gap widens, reducing the ratio of double-strip events as discussed in Section \ref{sec_ratio_eve}.

\subsubsection{Ratio of Sharing Energy between Adjacent Strips} \label{sec:beam_sub_exp}
To confirm whether sub-strip position resolution can be achieved, we investigated the variation in the sharing energy between adjacent strips as a function of the incident X-ray position for double-strip events.
The bottom panels of Fig.\,\ref{beam_cnt_ratio_14keV} and Fig.\,\ref{beam_cnt_ratio_22keV} show the relationship between the ratio of detected energies in adjacent strips, defined as $E_{ratio} = E_{i+1}/(E_{i} + E_{i+1})$, and the incident X-ray position $X_{pos}$ for the 14 keV and 22 keV beam scans. Here, $E_{i}$ and $E_{i+1}$ represent the energies detected in the adjacent $i$-th and ($i+1$)-th strips, respectively.

The ratio of sharing energies in adjacent strip $E_{ratio}$ varies continuously as the X-ray position moves from the $i$-th to the ($i+1$)-th strip on both the cathode and anode sides in the gap region, spanning approximately 30 $\mathrm{\mu m}$ and 50 $\mathrm{\mu m}$, respectively.
In addition, the properties of the sharing energies are also common across strips, with no significant individual differences, suggesting that a resolution smaller than the strip width can be achieved by using this relationship as demonstrated in the following.

\subsection{Development of Sub-strip Position Reconstruction Method}\label{sec:substrip_methode}
Based on the relationship found in Section \ref{sec:beam_sub_exp}, we developed a sub-strip position reconstruction method, focusing on data from the 14 keV beam on 80 $\mathrm{\mu m}$ strip-pitch region.

\begin{figure}[htb]
  \begin{center}
  \includegraphics[width=1\hsize]{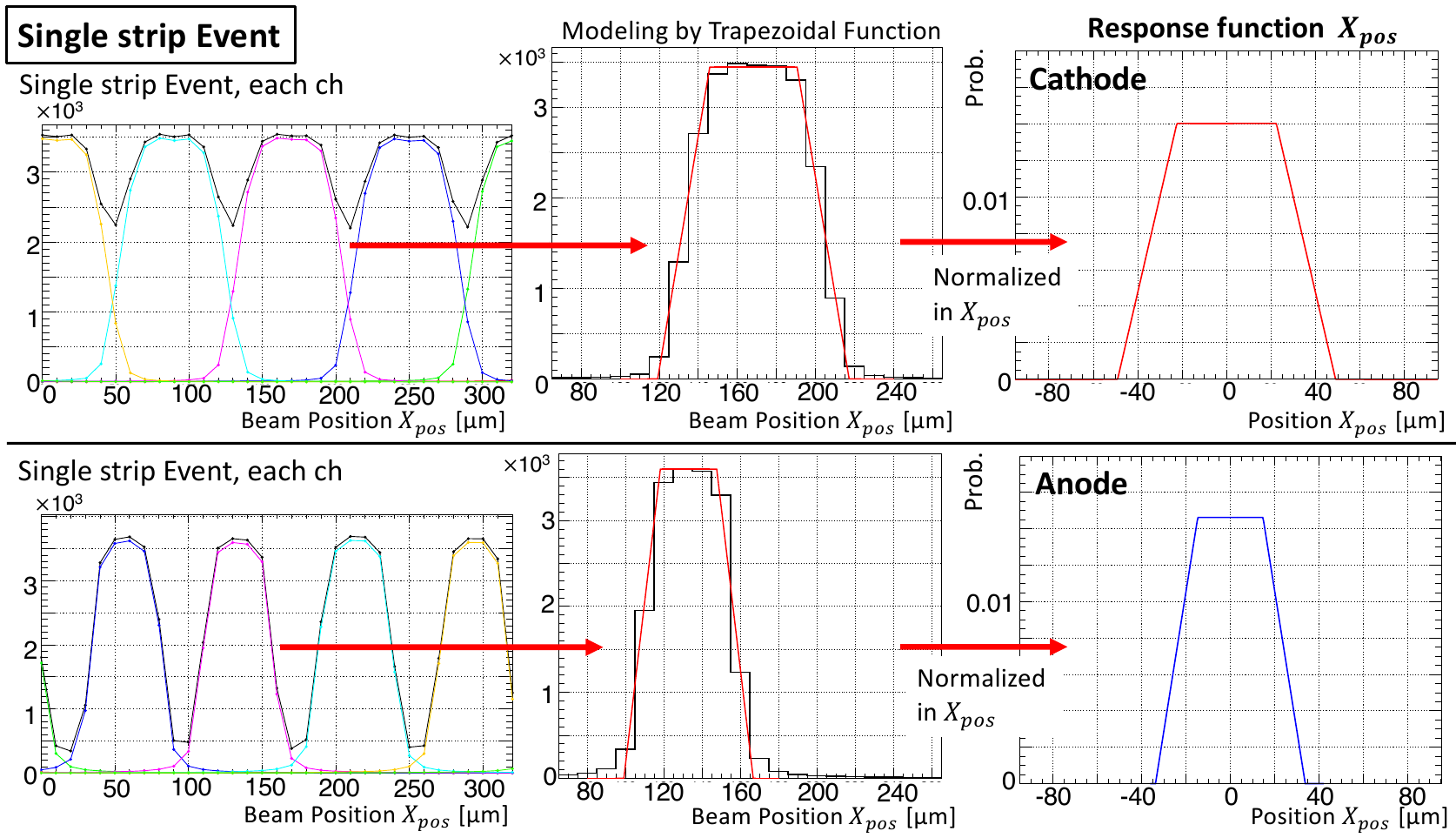}
  \caption{Procedures to create a response function to determine the sub-strip position $X_{pos}$ for single-strip events. Left and Center: The Count rate of single-strip events, fitted with a trapezoidal function. Right: The resulting response function for single-strip events, which determines the probability of incident X-ray position when a single-strip event is detected (14 keV, on 80 $\mathrm{\mu m}$ strip-pitch)}
  \label{beam_model_single}
  \end{center}
%   \end{figure}
% \begin{figure}[htb]
  \begin{center}
  \includegraphics[width=1\hsize]{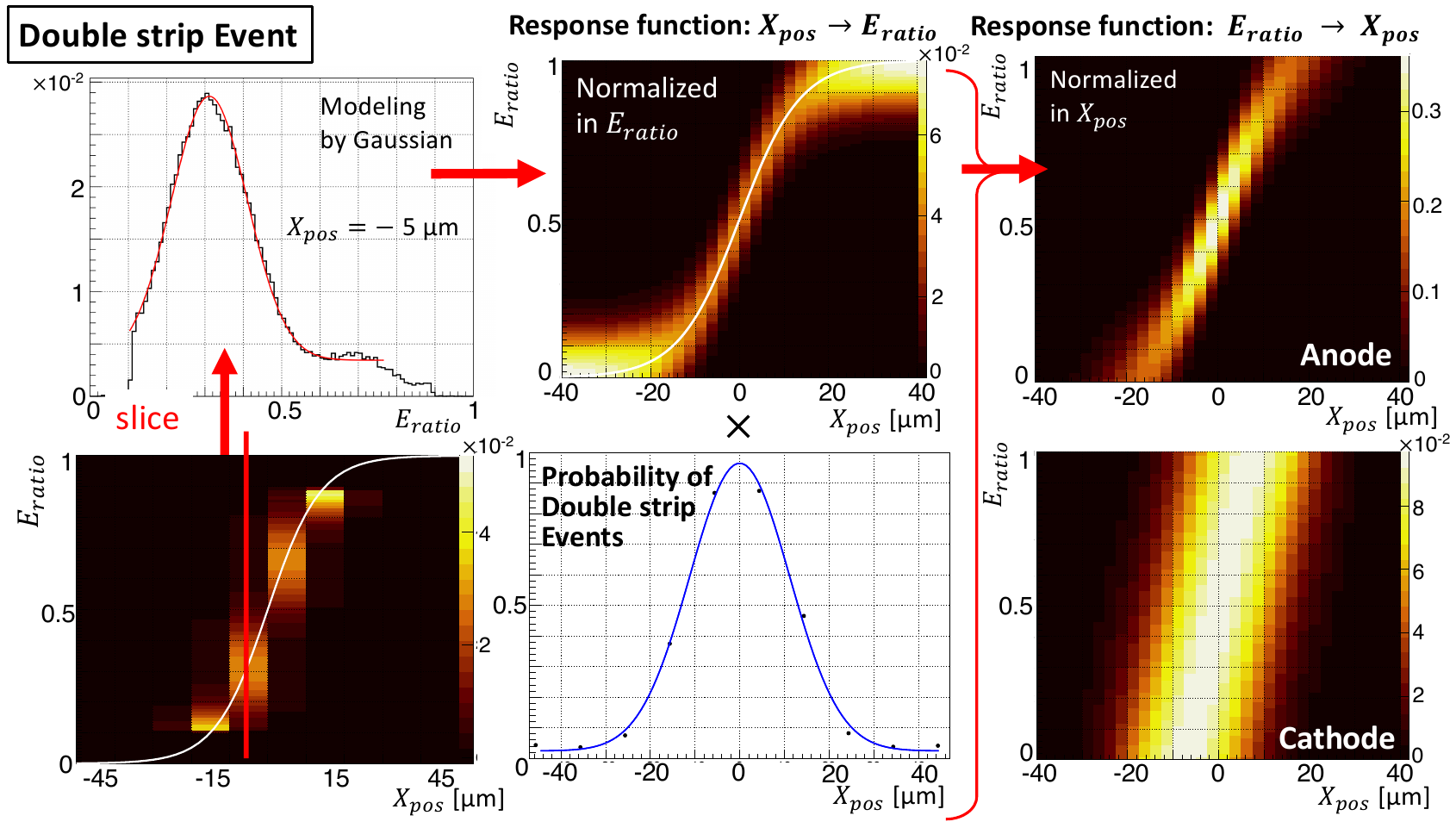}
  \caption{Procedures to create a response function to determine the sub-strip position $X_{pos}$ from the sharing energy ratio $E_{ratio}$ for double-strip events, based on beam scanning test results (14 keV beam on 80 $\mathrm{\mu m}$ strip-pitch region).}
  \label{beam_model_double}
  \end{center}
\end{figure}

For the single-strip events case, as shown in the left panels of Fig.\,\ref{beam_model_single}, the probability of detecting single-strip events is highest at the center of the strip region and linearly decreases toward the gap.
Therefore, we fitted the count rate of single-strip events with a trapezoidal function for both the cathode and anode side, as shown in the center panels of Fig.\,\ref{beam_model_single} center. By normalizing this fitted function across positions, we obtained a response function for single-strip events that estimates the probability of the incident X-ray position when a single-strip event is detected, as shown in the right panels of Fig.\,\ref{beam_model_single}.
On the anode side, the width of the top side of the trapezoid function is 38 $\mathrm{\mu m}$, comparable with the strip width. On the other hand, on the cathode side, the width is 52 $\mathrm{\mu m}$, slightly wider than the strip width due to charge loss in the gap regions, causing the signal on the adjacent strip to fall below the threshold.
This makes it difficult to determine the position with a precision less than 30 $\mathrm{\mu m} $ on the cathode side in the strip center region while it is still possible to achieve the sub-strip-pitch position resolution of approximately 50 $\mathrm{\mu m}$.

\begin{figure}[htb]
  \begin{center}
    \includegraphics[width=1\hsize]{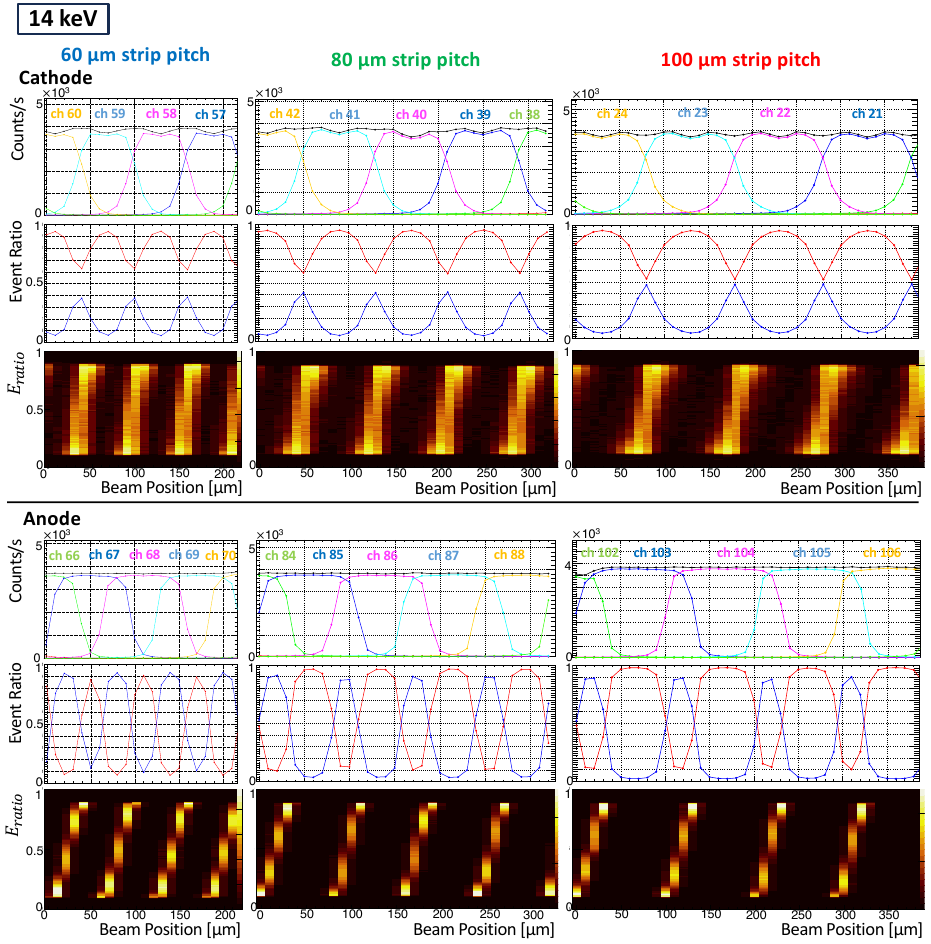}
  \caption{Beam scanning results for the 14 keV beam on the cathode and anode sides for each strip/gap width region. Top: Total count rates (Black) and individual count rates by channel (Colored). For double-strip events, one count is added to both channels. Middle: Ratio of single-strip (Red) and double-strip (Blue) events relative to the beam position. Bottom: Relationship between the detected energy ratio in adjacent strips $E_{ratio} = E_{i+1}/(E_{i} + E_{i+1})$ and the incident X-ray position $X_{pos}$.}
  \label{beam_cnt_ratio_14keV}
  \end{center}
\end{figure}

For the double-strip events case, we constructed a response function for sub-strip position resolution using the relationship between the incident X-ray position and sharing energy ratio, shown in the lower panel of Fig.\,\ref{beam_cnt_ratio_14keV}.
The distribution of $E_{ratio}$ at each beam position $X_{pos}$ is modeled by fitting a Gaussian function, with $\sigma \sim 0.1$ on the anode side and $\sigma \sim 0.4$ on the cathode side, and then normalized in each $X_{pos}$.
The centers of the Gaussian function across beam positions $X_{pos}$ are fitted with a sigmoid function, as shown in the white line in Fig.\,\ref{beam_model_double}.
This yields a response function representing the probability of sharing energy in adjacent strips for a given incident position, as shown in the top center panel of Fig.\,\ref{beam_model_double}.
To apply this function in position reconstruction, however, we need the response function to determine the incident X-ray position from the detected sharing energy in adjacent strips. 
Thus, we inverted this function by normalizing with respect to the beam position $X_{pos}$, weighting by the probability of double-strip events, as shown in the middle panel of Fig.\,\ref{beam_cnt_ratio_14keV}. 
This inversion produces a response function that determines the position from the detected sharing energy in adjacent strips, as shown in the right panels of Fig.\,\ref{beam_model_double}. 
On the anode side, this function enables precise position determination within $\sim 10$ $\mathrm{\mu m}$ resolution in the gap region. On the cathode side, while the probability distribution is slightly broader, the position can be determined in $\sim 20~\mathrm{\mu m}$ resolution in the gap region.

\begin{figure}[hbt]
  \begin{center}
  \includegraphics[width=0.7\hsize]{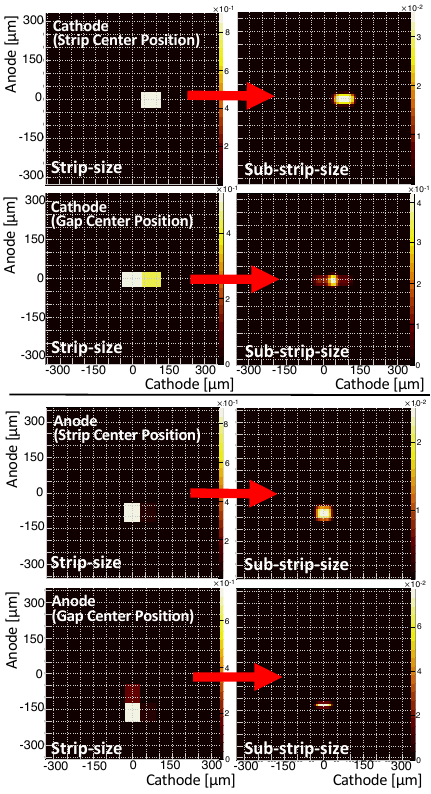}
  \caption{Strip-size image (pixel size: $60~\mathrm{\mu m} \times 80~\mathrm{\mu m}$) and Sub-strip image (pixel size: $10~\mathrm{\mu m} \times 10~\mathrm{\mu m}$) for the 14 keV beam scan on 80 $\mathrm{\mu m}$ strip-pitch region. The sub-strip image demonstrates enhanced position resolution beyond the intrinsic strip-pitch.}
  \label{beam_sub_image}
  \end{center}
\end{figure}

Fig.\,\ref{beam_sub_image} shows sub-strip images of the 14 keV beam scan on the 80 $\mathrm{\mu m}$ strip-pitch region, constructed using the response function for the sub-strip position reconstruction. 
When the incident position is near the gap center, where double-strip events dominate, position determination accuracy improves to approximately 20 $\mathrm{\mu m}$ on the cathode side and 10 $\mathrm{\mu m}$ on the anode side. 
In contrast, at the strip center, where single-strip events predominate, the position determination accuracy decreases to approximately 50 $\mathrm{\mu m}$ on the cathode side and 40 $\mathrm{\mu m}$ on the anode side. Nevertheless, incorporating information from the beam scan experiments allows the position resolution to surpass the intrinsic strip-pitch of 80 $\mathrm{\mu m}$.

\section{Discussion and Conclusion}
For the focal plane hard X-ray detectors of FOXSI-4, covering the energy range of 4 keV to 20 keV, an energy resolution less than 1 keV (FWHM) and position resolution less than 50 $\mathrm{\mu m}$ (FWHM, 5 arcsec) are required to spectrally and spatially resolve thermal and non-thermal emissions in solar flares. 
To achieve these requirements, we developed the wide-gap CdTe-DSD, incorporating a unique design principle to enhance position resolution by expanding the gaps between electrodes to induce charge-sharing across adjacent strip electrodes and using this sharing energy information for position reconstruction.
Beam scanning tests conducted at Spring-8 validated this concept, and the position of incident X-rays interacting near the gap center can be determined with an accuracy of 20 $\mathrm{\mu m}$. Even those interacting near the strip center can be determined with an accuracy of 50 $\mathrm{\mu m}$.
Although position resolution improved, wider gaps introduced challenges for energy resolution due to charge loss.
To mitigate this, we developed a new energy reconstruction method by fully utilizing both the cathode and anode sides of the energies and the adjacent strip energies, and the energy resolution (FWHM) of 0.75 keV at 13.9 keV is achieved.

The wide-gap CdTe-DSD for FOXSI-4 also allows us to evaluate how different strip and gap widths influence charge-sharing and position resolution. For enhancing charge-sharing to improve position resolution, we found that widening the gap to 30 $\mathrm{\mu m}$ with a strip width of 30 $\mathrm{\mu m}$ is sufficient for our purposes.
Further widening does not increase the region resolved with sub-strip position resolution on the anode side, although there is a minor increase on the cathode side, as discussed in Section \ref{sec:beam_ratio}.
Alternatively, widening the gap to 70 $\mathrm{\mu m}$ reduces the number of readout channels while expanding the effective area and maintaining energy resolution. However, as the gap widens, challenges arise in the low-energy band ($<10$ keV). Specifically, charge loss in the gap region can reduce signals below the trigger threshold, decreasing detection efficiency at the gap position, as discussed in Section \ref{sec:beam}.
In the case of solar hard X-ray observations with FOXSI-4, this limitation does not present an issue because the CdTe-DSDs are equipped with attenuators to reduce the flux and primarily detect photons above 10 keV. CMOS detectors cover the lower energy range, mitigating concerns in this band \cite{shimizu2024evaluation}. However, extending observations to X-rays below 10 keV, it becomes necessary to accurately model the differences in detection efficiency between the strip and gap regions and to reduce electronic noise so that triggers are generated even when X-rays enter the gap.

The energy and position reconstruction methods developed in this study performed well but were phenomenological. For future work, we plan to conduct simulations based on Monte Carlo methods and the Shockley-Ramo theorem to build a first-principles understanding of charge loss and charge-sharing due to wider gaps. This approach will support further optimization of both detector design and reconstruction algorithms, ensuring that the detectors not only meet but also exceed the requirements for future precise X-ray observations across a broad energy range.

\section*{Acknowledgement}
This work was supported by JSPS, Japan KAKENHI Grant Numbers 18H05457, 20H00153, 21H04486 and 22J12583.
The authors would like to thank the FOXSI-4 team, especially Hunter Kanniainen for developing the detector housing design, and Yixian Zhang, Kristopher Cooper, and Lindsay Glesener for the fruitful discussion on the detector response.
S.N. is also supported by Overseas Research Fellowship by JSPS and FoPM (WISE Program) and JSR Fellowship, The University of Tokyo, Japan.
Synchrotron beam experiments were performed at SPring-8 BL20B2 with the approval of the Japan Synchrotron Radiation Research Institute (JASRI) (Proposal No.\,2021B1542 and 2022B1477).

\bibliography{main}

% \begin{figure}[htb]
%   \begin{center}
%   \includegraphics[width=1.\hsize]{./fig/dsd_daq.pdf}
%   \caption{Pictures of the SpaceWire I/F board, the SPMU-001 FPGA board, and the Raspberry Pi4. All three boards can be stacked together, and the power can also be supplied directly from the SPMU-001 to Raspberry Pi4 and SpaceWire I/F board.}
%   \label{dsd_daq}
%   \end{center}
% \end{figure}



\end{document}